# DCcov: Repositioning of Drugs and Drug Combinations for SARS-CoV-2 Infected Lung through Constraint-Based Modelling

*Ali Kishk, Maria Pires Pacheco, Thomas Sauter**

University of Luxembourg
Faculty of Science, Technology and Medicine
6, avenue du Swing
L-4367 Belvaux
*Thomas.Sauter@uni.lu

## Abstract

The 2019 coronavirus disease (COVID-19) became a worldwide pandemic with currently no effective antiviral drug except treatments for symptomatic therapy. Flux balance analysis is an efficient method to analyze metabolic networks. It allows optimizing for a metabolic function and thus e.g., predicting the growth rate of a specific cell or the production rate of a metabolite of interest. Here flux balance analysis was applied on human lung cells infected with severe acute respiratory syndrome coronavirus 2 (SARS-CoV-2) to reposition metabolic drugs and drug combinations against the replication of the SARS-CoV-2 virus within the host tissue. Making use of expression data sets of infected lung tissue, genome-scale COVID-19-specific metabolic models were reconstructed. Then host-specific essential genes and gene-pairs were determined through *in-silico* knockouts that permit reducing the viral biomass production without affecting the host biomass. Key pathways that are associated with COVID-19 severity in lung tissue are related to oxidative stress, as well as ferroptosis, sphingolipid metabolism, cysteine metabolism, and fat digestion. By *in-silico* screening of FDA approved drugs on the putative disease-specific essential genes and gene-pairs, 45 drugs and 99 drug combinations were predicted as promising candidates for COVID-19 focused drug repositioning (https://github.com/sysbiolux/DCcov). Among the 45 drug candidates, six antiviral drugs were found and seven drugs that are being tested in clinical trials against COVID-19. Other drugs like gemcitabine, rosuvastatin and acetylcysteine, and drug combinations like azathioprine-pemetrexed might offer new chances for treating COVID-19.

### Introduction

Constraint-based modelling (CBM) approaches have successfully been applied in fundamental research [1], [2], [3] especially in cancer research [4], [5], [6], [7], as well as in microbial engineering [8], [9] among other research fields. CBM uses data- and prior knowledge-driven constraints to identify feasible metabolic flux distributions for a given condition [10]. Many communities and collaborative works contributed to reconstructing organism-specific generic metabolic networks which serve as starting points for CBM. Examples of such generic models are Recon 2 [11], Recon 2.2 [12], Recon3D [13], Human1 [14], and HMR [5]. Other types of metabolic models are context-specific models, that are built from tissue or disease-specific data. Usually, the context-specific models are draft reconstructions built from expression data of this condition by building algorithms like FASTCORE [15], rFASTCORMICS [7], INIT [5], and RegrEX [16] / or manually curated such as for *E. coli* [8], hepatocyte [17], and *Zea mays* [18]. These models are often used as scaffolds for the integration of omics data or more interesting to simulate the metabolic phenotypes of organisms, tissues, or cell lines.

Within the CBM methods, flux balance analysis (FBA) is a linear programming-based approach that maximizes or minimizes an objective function, often growth rate, to identify the optimal flux distribution(s) [19] [20]. *In-silico* knockout studies are common in FBA through gene or reaction deletion. This deletion may be single, double, or multiple [21]. The goal of single reaction deletion is finding the most critical reactions in respect to the objective function through brute force removal of each reaction individually and calculating the ratio of the objective rates between mutated and wild type models. Gene deletion studies are taking advantage of Boolean representations of the gene-reaction links known as Gene-Protein-Reaction (GPR) rules [22]. Gene deletion helps in defining essential genes whose deletion impacts the flux through the objective function [20]. Essential genes are often used as targets for drug repositioning.

The 2019 coronavirus disease (COVID-19) is caused by a strain of coronavirus called severe acute respiratory syndrome coronavirus 2 (SARS-CoV-2). COVID-19 was declared as a global pandemic on 11 March 2020 by WHO [23]. Human-to-human transmission is caused by droplets [24] and possibly aerosols [25]. Also, asymptomatic patients can transmit SARS-CoV-2 [26]. The virus might have originated from beta-coronaviruses from bats and pangolins [27]. SARS-CoV-2 can cause upper and lower respiratory infections, increasing its transmissibility and severity. SARS-CoV-2 utilizes the human protein angiotensin I converting enzyme 2 (ACE2) for cell entry with its spike protein. ACE2 is expressed on lung epithelial cells and other organs. ACE2 converts Angiotensin II (AT-II) to Angiotensin-(1-7) (AT-1,7) negating the inflammatory effect of AT-II [28]. Thus, SARS-CoV-2 infection decreases the concentration of cellular unbound ACE2 molecules to facilitate the cell entry, causing an increase of AT-II which eventually increases the oxidative stress ion superoxide [29]. SARS-CoV-2 have shown a relatively low mutation rate, lower than SAR-CoV-1, despite worldwide transmission [30]. Also, the spike protein is more conserved than other proteins [30]. People with increased COVID-19 risk are patients with cancer [31], chronic kidney disease [32], obesity [33], type 2 diabetes mellitus [34], immunocompromised [35], cardiac diseases [36], COPD [36], and sickle cell disease [37].

Acute respiratory distress syndrome (ARDS) is one of the severe symptoms of COVID-19, which may be attributed to alveolar epithelial cell injury [38]. COVID-19 induced ARDS lasts 8-12 days, which exceeds the known ARDS onset of one week [38]. This ARDS may become unresponsive to invasive mechanical ventilation and increase lung injury [39]. Severe COVID-19 courses are also associated with acute injury to heart, kidney, and cerebrovascular diseases [40]. A systematic meta-analysis of 27 histological studies found that COVID-19 affects in addition to the previous organs, vasculature, central nervous system, and hemolymphatic system [41]. Despite the COVID-19 pandemic is still in its few first months, long term effects for COVID-19 survivors (or long-haulers) have become to emerge such as new-onset diabetes, increasing severe complications in pre-existing diabetes [42], fatigue, dyspnea, psychological distress [43] and myocardial inflammation [44]. The cause of these long-term effects is still unclear whether the result of COVID-19 or the treatment, also these symptoms were reported from relatively small populations.

Metabolic modeling and in particular FBA were often used to understand the effects of microbes on human cells. Notably, a human alveolar model was used to assess the metabolic interaction between the host and M. tuberculosis [45]. Lately, a similar approach was applied in the viral genomes to model the impact of the Chikungunya, Dengue, and Zika viruses on macrophage [46].

Only a few studies employed FBA on COVID-19 so far. *Renz et al.* used the viral genome information available at that time, to generate a SARS-CoV-2 specific viral biomass objective function (VBOF) [47]. This VBOF generation from the genome information consisted of six steps on nucleotide investment, amino acid investment, ATP requirements, pyrophosphate liberation, total viral molar mass, and final construction of the VBOF [46]. Then the VBOF was added to a human alveolar macrophage model (*iAB-AMØ1410*) [45] to build an infected model. They identified guanylate kinase (GUK1) as an essential gene through *in silico* knockout, that allows decreasing the viral biomass without affecting the human biomass maintenance. Several worldwide collaborative works in computational modeling of COVID-19 was established. *Ostaszewski et al.* built the COVID-19 Disease Map to understand the mechanistic interactions between SARS-CoV-2 and human tissues [48]. In another collaborative study, *Gysi et al.* applied network analysis for drug repositioning using three different ranking approaches: network proximity, diffusion, and deep learning-based [49]. To further support the search for an effective treatment for COVID-19, we employ here FBA to find candidate drugs and drug combinations that target viral-specific essential genes in COVID-19 infected lung cells through context-specific models built from expression data by the rFASTCORMICS workflow [7] (see **Fig. 1**). We also highlight key pathways of these essential genes that might contribute to COVID-19 severity.

# Methods

# A. SARS-CoV-2 Essentiality Analysis in Lung

## A.1 Differentially Expressed Genes Analysis

### *A.1.1 Data Preprocessing*

At the onset of the pandemic, two datasets were available focusing mainly on the effects of the virus on lung tissues. These two bulk RNA seq datasets GSE147507 [50] and GSE148729 [51] of human cell lines hosting SARS-CoV-2, as well as of mock samples were downloaded from the NCBI Gene Expression Omnibus (GEO) [52] data repository on April 23, and May 15, 2020, respectively. The GSE147507 dataset, which focuses on the expression changes at various severity levels of Infection (severity study), contains 36 samples originating from healthy epithelial, A549, and Calu-3 cells infected by SARS-CoV-2 at three different viral loads, as well as control samples with a mock infection. Also, the samples of two conditions were transfected with a vector expressing ACE2 (**Table 1**), a protein that governs SARS-CoV-2 cell entry in the host cells. Conditions with two replicates only or subjected to drug perturbations were not considered for the analysis. Raw counts were converted to the Reads Per Kilobase of transcript (RPKM) using an in-house Python script.

For the GSE148729 dataset, which monitors how expression changes at different points after the infection (time-series study), the normalized Fragments Per Kilobase of transcript per Million (FPKM) values of Calu-3 and H1299 cell lines infected by SARS-CoV-2, as well as controls were retrieved from GEO.

**Table 1: Severity Study Metadata (GSE147507).** Expression data from three lung cell lines infected with SARS-CoV-2 at three different viral loads, and for some samples transfected with a vector expressing ACE2 with their controls.

| Condition | Cell Line | Multiplicity of infection | ACE2 Vector | Abbreviation | Number of Samples Infected/ Mock |
|---|---|---|---|---|---|
| Series 1 | NHBE | 2 | No | NBHE_2 | 3/3 |
| Series 2 | A549 | 0.02 | No | A549_0.02 | 3/3 |
| Series 5 | A549 | 2 | No | A549_2 | 3/3 |
| Series 6 | A549 | 0.2 | Yes | A549_0.2_ACE2 | 3/3 |
| Series 7 | Calu-3 | 2 | No | Calu3_2 | 3/3 |
| Series 16 | A549 | 2 | Yes | A549_2_ACE2 | 3/3 |

**Table 2: Time-series Study Metadata (GSE148729).** Time series expression data with five time-points (4, 8, 12, 24, and 36 hours) with infected and mock samples for two lung cell lines.

| Condition | Cell Line | Time Point in hrs | Number of Samples Infected/Mock |
|---|---|---|---|
| Calu3_4h | Calu-3 | 4 | 4/4 |
| Calu3_8h | Calu-3 | 8 | 2/0 |
| Calu3_12h | Calu-3 | 12 | 4/2 |
| Calu3_24h | Calu-3 | 24 | 2/2 |
| H1299_4h | H1299 | 4 | 2/2 |
| H1299_12h | H1299 | 12 | 2/0 |
| 1299_24h | H1299 | 24 | 2/0 |
| H1299_36h | H1299 | 36 | 2/2 |

### *A.1.2 Differentially Expressed Genes Analysis:*

For the severity study, a principal component analysis was first performed using FactoMineR [53] to identify and if necessary, remove outliers by visual inspection. Then genes with low expression values were filtered out using edgeR (Version 3.30.3) 's *filterByExpr* function. This function keeps genes based on a minimum count-per-million in at least k samples, determined by the lowest sample size between all conditions [54]. Finally, *DESeq2* (Version 1.28.1) [55] was run on the preprocessed data to identify differentially expressed genes (DEG) between the infected and the mock samples applying an adjusted P-value threshold of 0.05, and an absolute log fold change threshold of 1. edgeR and DESeq2 were used via R (Version 4.0.1),

### *A.1.3 Metabolic Pathway Analysis of the DEGs*

To identify metabolic alterations in the metabolism of infected lung cells, for each condition, DEGs were mapped to the genes of the generic model Recon3D_01 [13] via the GPR rules to retrieve differentially expressed reactions (DERs) as well as their associated pathways.
As condition NHBE_2 and A549_0.02 of the Severity Study only displayed a few DEGs, these conditions were not included in the downstream analysis. For each pathway with at least three reactions, the number of up- and down-regulated reactions were summed up, and the reactions per pathways for each condition and pathway was computed as follows: The enrichment overlap is the number of shared reactions between a condition and a pathway, divided by the total number of reactions in this pathway. To improve the readability of the plot, only pathways with more than five percent of DERs were depicted.

### A.2 Essentiality Analysis

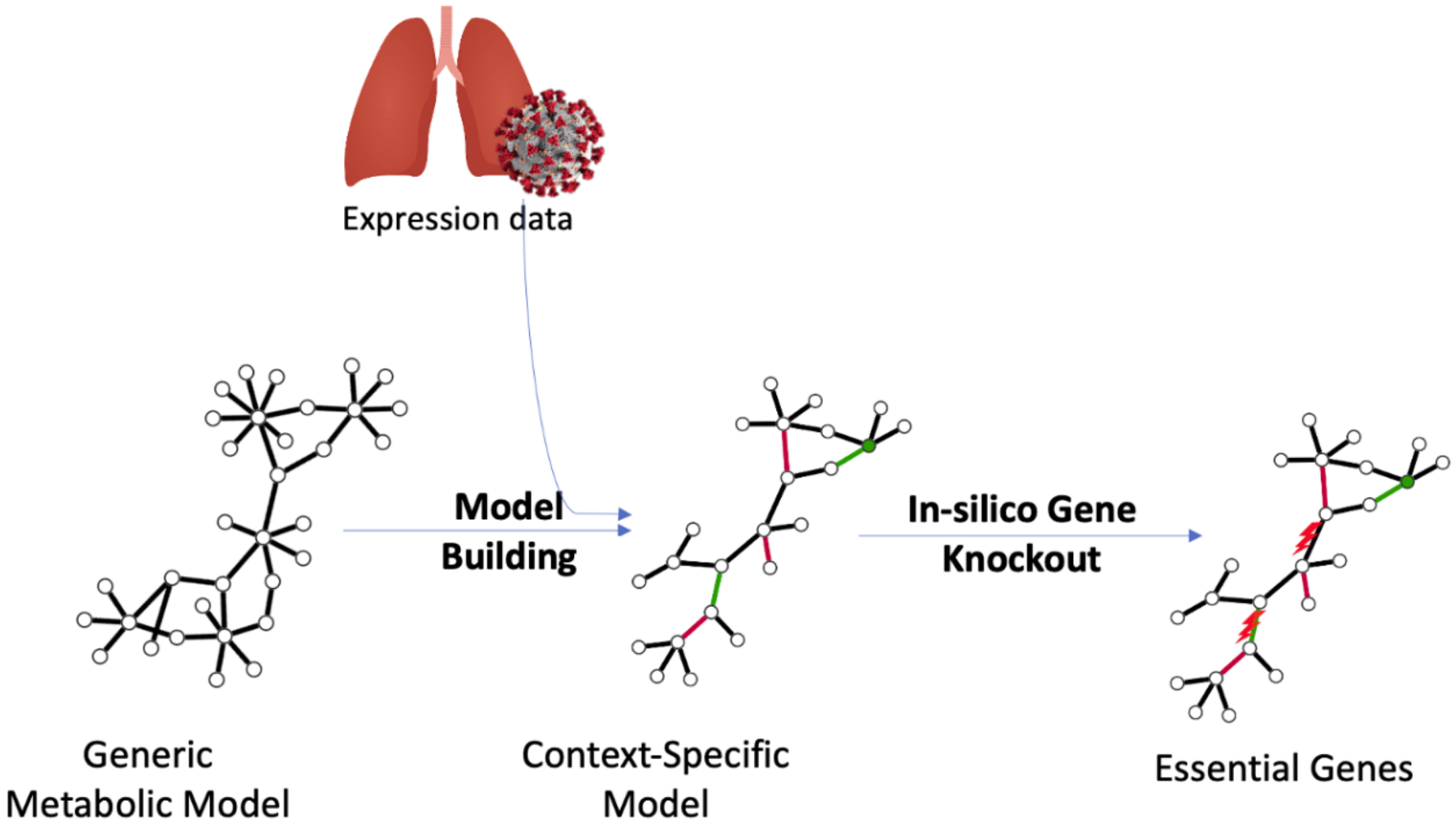

**Fig. 1: Overview of the pipeline of essential gene prediction for COVID-19 infected lung cells.** The viral biomass function (VBOF) was added to the generic models as detailed in *(A.2.1)* to build the infected generic model. Mock and infected lung expression data were used to build the context-specific models using rFASTCORMICS and a consistent version of Recon2.04 as input for control models and the consistent version of Recon2.04 with the added viral biomass for infected models, respectively. The objective functions were adjusted as explained in *(A.2.2)*. Finally, essential genes were identified by *in-silico* gene knockout.

### *A.2.1 Condition-specific Model Building*

To further elucidate, the metabolic alteration provoked by the virus, metabolic models for the infected samples and mock samples were built. Therefore, the VBOF from the infected alveolar macrophage model *iAB_AMO1410_SARS-CoV-2* (MODEL2003020001) [47] was added to the generic reconstructions Recon2.04 [11] and Recon3D_01 [13] using the *addReaction* function of the COBRA Toolbox [20]. The identifiers of metabolites included in the biomass had first to be modified to match the ones of the generic reconstruction. Then, FASTCC [15] was run to remove blocked reactions. For each condition, the RPKM values and the modified consistent generic reconstruction were used as input for the rFASTCORMICS [7] to get condition-specific models. COBRA Toolbox and FASTCC were used via MATLAB (R2019a).

### *A.2.2: Single Gene Knockout*

The metabolic models were then used to identify viral-specific vulnerabilities, using a single gene deletion approach on the mock and infected models. For the infected model, to ensure that both host and viral biomasses objective functions can carry simultaneously a flux in the infected models, the objective coefficients were set to 100 and 1, respectively, and the upper bound of the host biomass was fixed to 10% of its maximal flux. This setting constrains the model to guarantee cell homeostasis and protein turnover in the host model while diverting all non-essential resources for viral reproduction.

model.c(viral biomass) =1
model.c(host biomass) =100
model.ub(host biomass) = 10 % of max flux determined by FBA

*In-silico* single-gene knockouts (SKO) were performed on the infected models using a corrected version of the *singleGeneDeletion* function of the COBRA ToolBox [20] to assess the impact of the knockout of each gene on the viral biomasses. The 0.2 threshold was used as a cutoff for gene growth rate Ratio (grRatio) to identify essential genes.

### *A.2.3 Double Gene Knockout*

To identify potential targets for drug combinations, double gene knockouts (DKO) for all gene pair combinations were simulated using the *doubleGeneDeletion* function on the infected models. From the analysis, we obtained two lists of synergistic gene pairs: non-essential gene pairs that allow reducing the growth rate below the specified thresholds when simultaneously knocked-out and pairs of essential and non-essential genes that induced a stronger reduction of the growth than the knockout of the essential gene alone. Both non-essential and essential and non-essential gene pairs were concatenated as DKO outputs for further drug repositioning.

### *A.2.4 Essentiality and Safety Scoring*

To test the knockout impact of SARS-CoV-2- infected host-specific predicted essential genes and gene pairs on the healthy counterpart tissue, SKO, and DKO of these genes were performed on the healthy models. Genes or gene pairs that cause a reduction of biomass only in the infected models are considered safe, whereas those that cause also a reduction of biomass in the healthy models are regarded as potentially toxic. The essentiality score of an essential gene is the sum of infected models that show this gene as essential. The safety score of an essential gene is the sum of healthy models that show this gene as safe. A healthy model SKO and DKO of a gene was applied only if the gene is determined as essential in its respective infected model. A scatterplot of essentiality scores against safety scores was plotted (see **Fig. 3**).

## B. Gene Enrichment of the potential targets

The identifiers of the essential genes and synergetic genes were translated into HGNC gene symbols using *GSEApy* (Version 0.9.17, https://github.com/zqfang/GSEApy/) Python package (3.7.4) and then uploaded to Enrichr API [56] to identify enrichment of these genes in KEGG pathways (KEGG2019 human [57] database with 0.05 P-value cutoff). All enrichment results based on only one gene were discarded. Then enrichment percentage was calculated as in (***A.1.3).*** Metabolic pathway analysis was also applied using the Recon3D_01 *subSystem* as background instead of KEGG pathways on the essential genes such as in (A.1.3) without further filtering of the pathways. Comparison of pathways of Calu-3 and NHBE cell lines were excluded from the analysis of the effect of viral load and ACE2 vector in the severity study, since these cell lines didn't have ACE2 vector condition.

## C. Drug repositioning of the Essential Genes

To identify drugs targeting the predicted viral-specific essential genes, drug-target interactions were downloaded from DrugBank [58] on April 23, 2020. Drugs that were withdrawn, nutraceutical or experimental were discarded from the analysis. Drugs that are described as having any effect on the potential targets were selected as candidate drugs (**Table S3**) and drug combinations (**Table S4, S5**). To determine which drugs, have multi-target effect, tripartite networks of the drug-gene-pathway interactions were constructed for the single and double knockout drugs using Recon3D_01 subSystems as pathways (**Fig. 3. b., Fig. 5**). The tripartite networks were constructed using the *sankeyNetwork* function in networkD3 [59, p. 3] (version 0.4) package in R.

## D. Relationship with Ferroptosis

As SKO targets were enriched for many pathways related to ferroptosis, the potential targets and SKO drugs were searched in a curated database (FerrDb) [60] for ferroptosis genes, and related drugs. FerrDb classifies genes into driver, suppressor, and marker, while it classifies drugs into inducer and inhibitor. These classes were also used to identify the role of the potential targets and SKO drugs in the ferroptosis pathway.

# Results

The main goal of the present study is to understand metabolic changes induced by COVID-19 in several lung cell lines, at various severity of infection, and at different time points after the infection. We then used single and double knockouts to identify vulnerabilities that are specific to infected cells that are predicted by our network models to reduce the viral proliferation while only affecting moderately the growth of host and control cells. To further prioritize essential genes, we considered their essentiality scores across cell lines and in time and the effect of a knockout of these genes in the healthy tissues. The final aim is to identify conserved essential genes across infected models that do not provoke severe side effects when the gene is knocked-out in the healthy counterpart model. To further identify vulnerabilities in the networks that could be exploited as drug targets, *in silico* inactivation of reactions were simulated. We further investigated the pathways harboring the predicted essential genes and reactions, to gain insight into how the virus adapts to the metabolism of lung cells. Finally, we proposed drug and drug combinations that target the predicted essential genes and synergistic essential gene pairs.

### Metabolic Pathway Analysis of Differentially Expressed Genes indicates COVID-19 based rewiring of core metabolism

Infection by the SARS-CoV-2 virus provokes alterations of the metabolism of the host cells. To elucidate these induced metabolic changes, we took advantage of two available expression data sets (Severity Study; Time-series Study; see Methods for details). PCA of the severity study samples, shows a clear cluster separation according to the cell type (see **Fig. S1**). Besides determining the differentially expressed genes (DEG), we built genome-scale metabolic models applying rFASTCORMICS (see Methods, ***A.2.1***). This resulted in 50 models (28 infected and 22 mock) with a median of 3646 metabolites (2465-5088) and 2456,5 reactions (1790-3474). To determine the key dysregulated pathways, we mapped the DEGs on the RECON3D_01 model and displayed the pathway alterations in the mostly dysregulated conditions (**Fig. 2**, and **Fig. S2** that shows a representation of all pathways without filtering on the number of reactions, nor the reactions per pathway). A549_0.02 condition didn't show any differentially expressed metabolic genes, thus it was discarded from the DEGs metabolic pathways. Meanwhile NHBE_2 condition pathways were filtered (see Methods, ***A.1.3***). Among the most down-regulated pathways in the A549 cell lines with transfection (A549_2_ACE2 and A549_0.2_ACE2) in comparison to no ACE2 vector (A549_2) were chondroitin sulfate degradation, phosphatidylinositol phosphate metabolism, and phenylalanine metabolism, whereas glutathione metabolism was upregulated. For the A549 cell line with high viral load (A549_2_ACE2 and A549_2) in comparison to low viral load (A549_0.2_ACE2), a downregulation of fatty acid synthesis, androgen and estrogen synthesis and metabolism, chondroitin synthesis, and pyruvate metabolism were also detected. Across all conditions including Calu3_2, we found a moderate downregulation of several pathways (glycerophospholipid metabolism, glycosphingolipid metabolism, sphingolipid metabolism). Other regulated pathways in Calu3_2 are the downregulated chondroitin sulfate degradation, nucleotide interconversion, and the upregulated cholesterol metabolism.

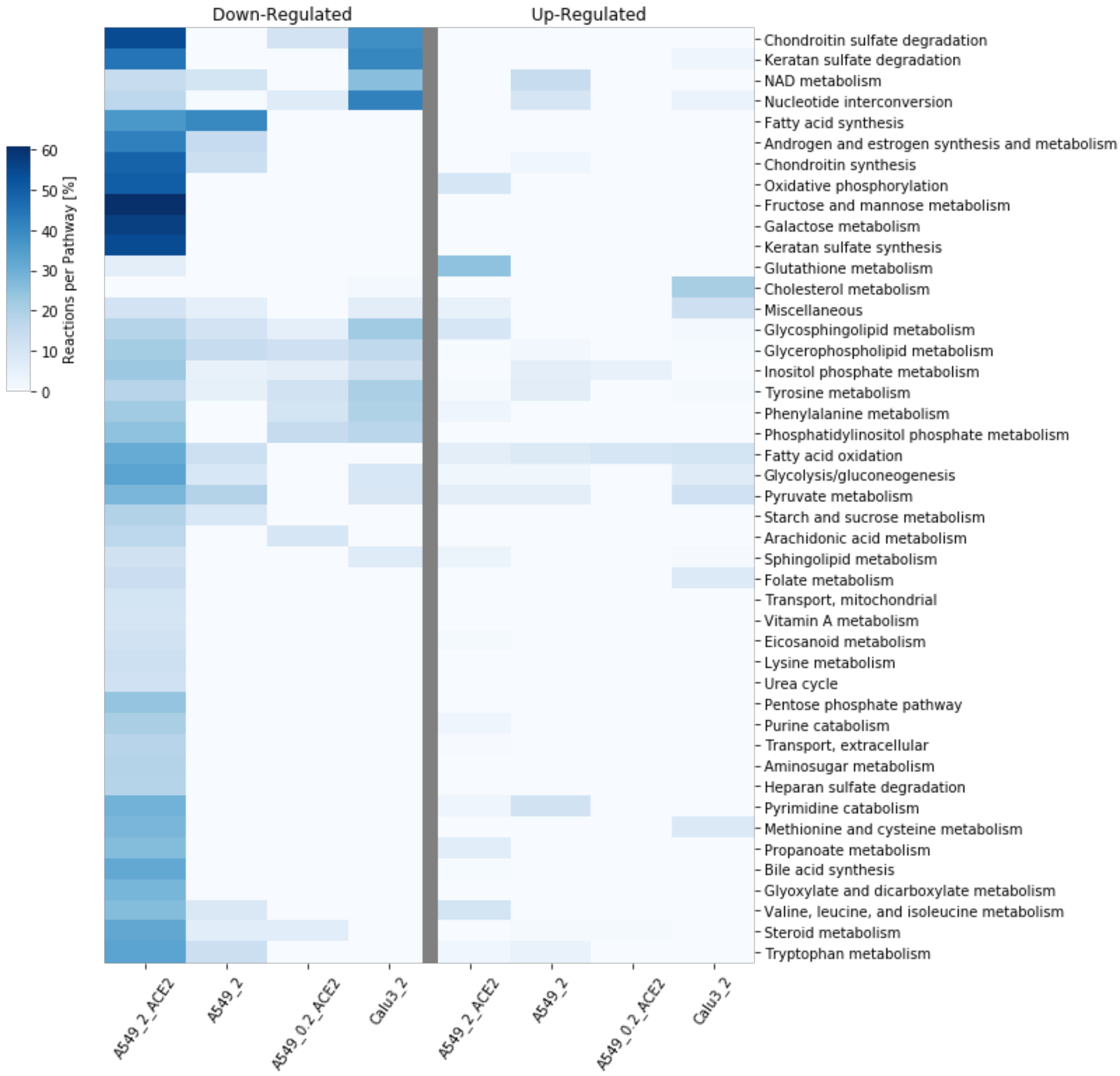

**Fig. 2: Reactions per Pathway Heatmap for Pathway Analysis of Differentially expressed genes in the Severity Study.**

Differentially expressed genes (DEGs) were computed with DESeq2. The Down- and up-regulated were mapped to the pathways (subSystems) of Recon3D_01. The number of up and down-regulated reactions was then summed up to identify the top altered pathways in the infected lung cell lines in the severity study (A.1.3). The color code "Reactions per Pathway [%]" represents the number of enriched metabolic reactions in a pathway divided by the overall number of reactions in this pathway. The transfection of ACE2 at an MOI of two in the A549 cell lines, caused the downregulation of many pathways which was not seen when the MOI was decreased by a factor ten.

## *In silico* single gene deletion predicts common potential drug targets across infected cell lines with reduced side effects on control cells

The DEGs and the performed pathway analysis indicate a rewiring of the metabolism induced by SARS-CoV-2. The next step was then to verify if these alterations caused the appearance of infected cell-specific essential genes that could specifically be targeted by repositioned drugs. Therefore, for every condition of both lung studies, *in silico* single gene deletion was performed on the respective reconstructed metabolic models. 23 unique genes were predicted to be essential in the infected models (**Table S1,4**). To assess if a drug targeting the candidate essential gene will kill the infected cell or reduce the viral proliferation, we computed an essentiality score for each essential gene, which sums up the number of models in which this gene is predicted to be essential. Essential genes that are only found in one or a few conditions might be cell line or experiment (e.g., medium) specific and hence might not have general biological relevance. Single gene deletion of each predicted essential gene was then performed on the counterpart control model (see Methods, A.2.2), to predict the effect of the gene knockout on the healthy tissue. This allowed obtaining a safety score and hence estimating the potential toxicity of each of the considered drug targets.

The obtained essentiality and safety scores are plotted for visual inspection (see **Fig. 3**). CRLS1 and SGMS1 scored highest for essentiality, but were among the lowest for safety, thus indicating that targeting any of these genes might be effective against the virus but also reduces the growth of healthy cells suggesting high toxicity of respective drugs. On the other hand, GUK1 gene showed a moderate essentiality score, but a higher safety score expecting fewer side effects. No gene could be identified that has high efficiency and safety. Also, nine of the 23 essential genes belong to the SLC transporter gene family. Transporters are known key regulators of metabolic flux [7], [61], hence modulating their expression might contribute to diverting metabolic fluxes to pathways for viral survival and proliferation. Of the 23 essential genes, 10 genes were shared between the two investigated lung studies, and many essential gene sets are shared between the investigated conditions (CRLS1, GUK1, SGMS1 in the severity study and CRLS1, ISYNA1, SGMS1, SLC27A4 in the time-series study), suggesting the existence of a consistent metabolic rewiring of the host metabolism rather than random alterations.

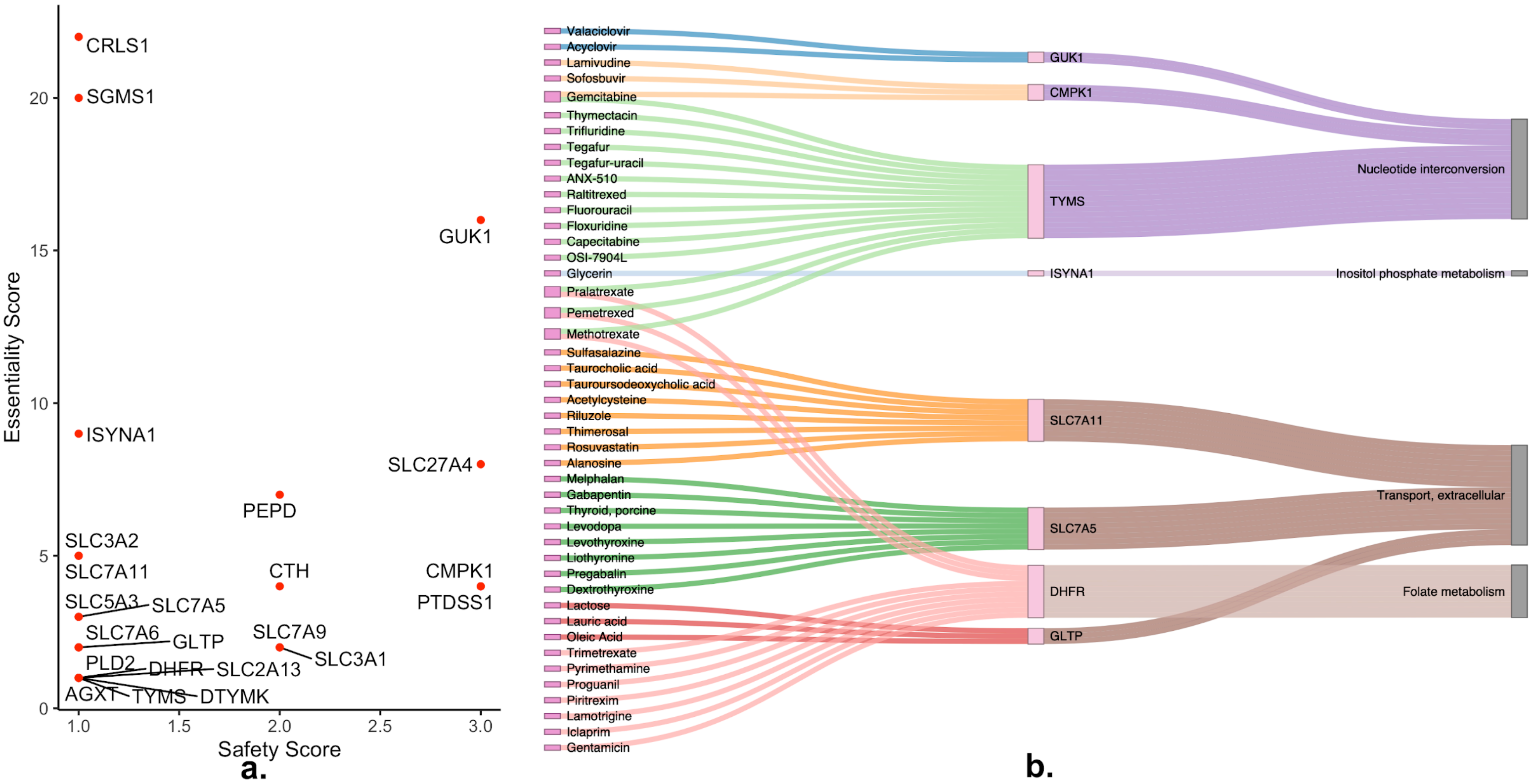

**Fig. 3: Scatterplot and Tripartite Network of Essential Genes, and their Predicted Drugs and Pathways, Determined by *in silico* Single-gene Deletions on the Infected Lung Models. a. Scatterplot of essentiality and safety scores of the essential genes.** Essentiality and safety scores correspond to the number of infected and healthy models respectively,

in which each gene is predicted to be essential. The y-axis indicates the number of infected cell lines for which the gene is predicted to be essential, whereas the x-axis indicates the number of control cell lines that are predicted to remain unharmed by the silencing of the target genes. **b. Tripartite network of the drug-gene-pathway interactions of the essential genes:** A network of the single repositioned drugs and their essential genes, predicted by *in silico* gene deletion, was built. The relationships between the essential genes and their pathways were mapped using Recon3D_01 subSystem. Genes, and their connected drugs, that don't have pathways in Recon3D_01 subSystems were discarded.

### ***In silico*** **single gene deletion predicts potential drug targets for different stages and disease severity levels**

Variability in the metabolism of cell lines, viral load, and time of infection gives rise to the appearance of context-specific essential genes. Core essential genes in infected cells are optimal drug targets as likely to be efficient for a majority of patients. Essential genes that are specific to the time of the disease or severity level are also of interest, as it allows modulating the specific treatment. It might be reasonable to provide drugs with strong adverse effects to more severe cases and to opt for lighter treatments for mild affections. To identify core and context-specific essential genes, we performed *in silico* gene knockouts on all reconstructed models and compared the sets of essential genes across all the conditions and between the severity and time-series studies. And more specifically we focused on the effect of the transfection of the ACE2 vector, of the viral load in the severity study, and the time after infection in the time-series study. ACE2 is crucial for SARS-CoV-2 cell entry by binding with its spike protein, but ACE2 also has many cellular functions crucial to the host cells such as in the angiotensin-renin system. By comparing the essential genes in the absence (A549_2 &) and presence of the ACE2 vector (A549_2_ACE2 & A549_0.2_ACE2), we could identify one set of genes (ISYNA1, SLC3A2, SLC7A11) that are essential for the virus in the absence of the ACE2 vector. By comparing the essential genes in the A549 cell line in the severity study with a high multiplicity of infection (MOI) (A549_2 & A549_2_ACE2) against low MOI (A549_0.02 & A549_0.2_ACE2), we identified essential genes for high viral load (CMPK1, CTH, PTDSS1, SLC2A13, SLC3A1, SLC5A3, SLC7A9) in A549_2_ACE2 and one essential gene, DTYMK, in A549_2, where the gene set (AGXT, DHFR, SLC27A4, TYMS) were unique for low viral load in A549_0.2_ACE2.

For the time-series study, a list of core essential genes (7-10 genes) was common to every time point and for each cell line (see **Table S2**). Besides the core essential genes, there were time-point specific essential genes that were essential only at very specific time-points due to the inactivation of alternative pathways (**Table S2**). The Calu-3 cell line has eight core essential genes (CRLS1, GUK1, ISYNA1, PEPD, SGMS1, SLC27A4, SLC3A2, SLC7A11) and three time-point specific genes (see **Table S2**). Out of the eight core essential genes, five (SLC27A4, CRLS1, GUK1, PEPD, SGMS1) were also in the six essential genes of the Calu3_2 condition in the severity study. six core essential genes (CRLS1, ISYNA1, PLD2, SGMS1, SLC27A4, SLC7A6) and two time-specific genes were predicted for the H1299 cell line. Jaccard similarity of the essential genes between different conditions, shows cell type specific essential genes in the time-series study in (see **Fig. S4**). Clustering of the reconstructed models by core reactions using Jaccard similarity (see **Fig. S3**), shows four clusters by cell type (A549, H1299, NHBE, Calu-3), even for cell lines between the two studies (Calu-3). This cell line-specific clustering is more apparent in Recon 3D than in Recon2. Moreover, the infection state (Mock, Infected) forms subclusters within each of the four main clusters.

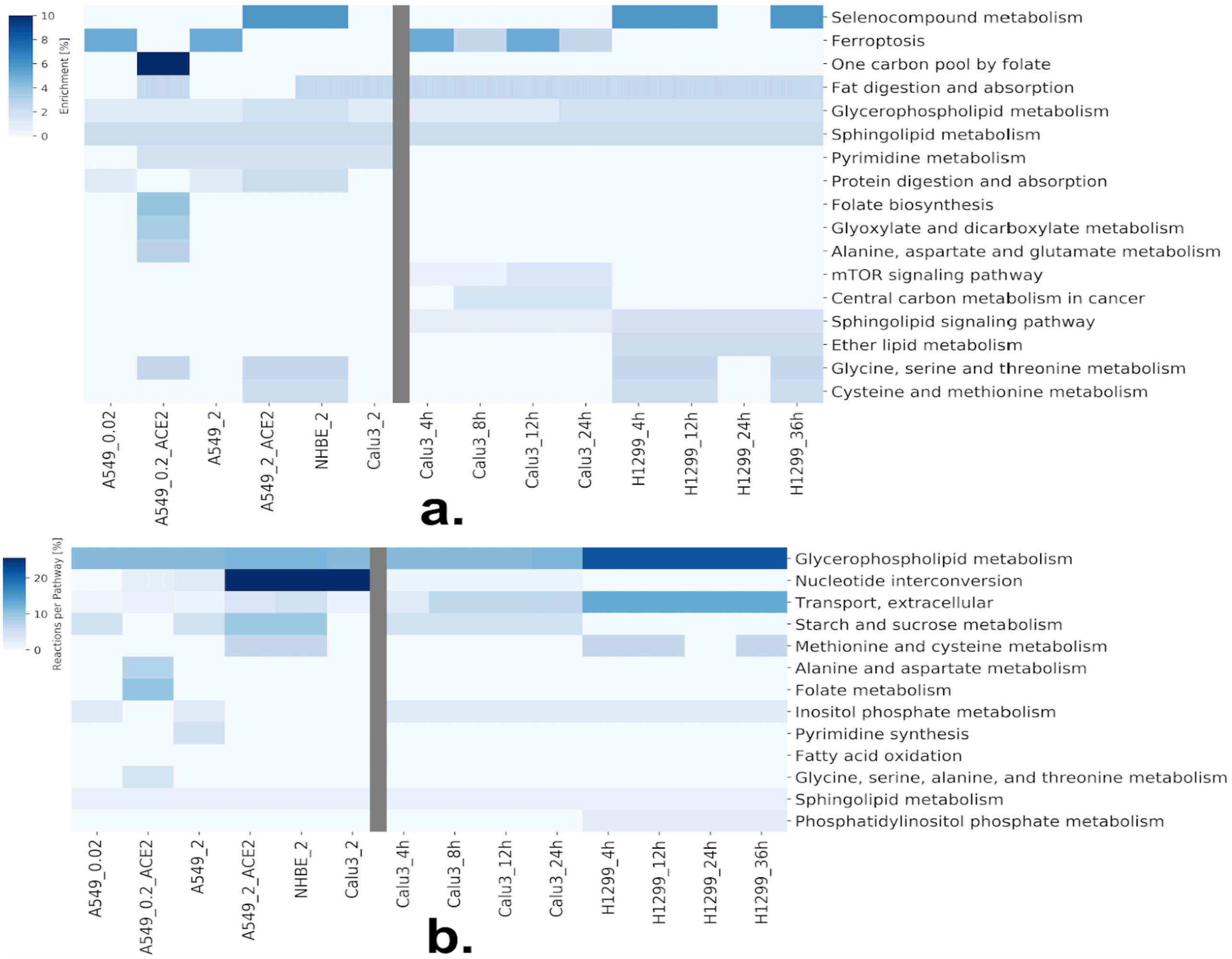

**Fig. 4: Pathway Analysis of the Essential Genes in the two Lung Studies.**
Identification of the pathways harboring the identified essential genes using Enrichr and Recon3D_01 subSystem, and hence the most critical pathways for the viral survival and proliferation across cell type, severity level, and time after infection.
a) Enrichr enrichment of the severity and time-series lung studies respectively as mentioned in (B). Conditions in the severity study were named as: cell line + ACE2 vector (if exists) + MOI. Conditions in the 2nd lung cell line were names as: cell line + time point. The color code in a) "Enrichment [%]" represents the number of enriched genes in this pathway divided by the overall number of genes in this pathway.
b) Metabolic pathway analysis of severity study and time-series lung studies respectively as mentioned in (B). The color code in b) "Reactions per Pathway [%]" represents the number of enriched metabolic reactions in a pathway divided by the overall number of reactions in this pathway.

## Essential genes and reactions are predicted to be harbored in 10 unique pathways among which is methionine and cysteine metabolism

To obtain a comprehensive picture of viral essentiality, we apply pathway analysis for core and context-specific essential genes to identify pathways that are major players in the determination of the severity as well as the stage of the infections. For both studies, the predicted essential genes were enriched in 13 unique pathways, of which eight were shared between both studies (**Fig. 4 b**). Glycerophospholipid metabolism was enriched in all conditions across both studies. Nucleotide interconversion and extracellular transport were highly enriched in the severity and in the time-series study, respectively. Also, two pathways were shared with the DEGs pathways (glycerophospholipid metabolism, sphingolipid metabolism). The essential gene ISYNA1, encoding a synthase in the inositol phosphate metabolism pathway, was specific to cell lines without ACE2 vectors. No unique change in the set of essential genes' pathways was found in the function of the viral load in both conditions (A549_2 & A549_2_ACE2).

To also explore pathways harbouring essential genes that are not directly linked to metabolism or that are not captured by the metabolic genes and pathways in Recon3D_01, an Enrichr pathway analysis was performed [56]. Among others, ferroptosis, selenocompound metabolism, cysteine, and methionine metabolism, mTOR signaling pathway, and ether lipid metabolism were enriched for the essential genes (**Fig. 4 a**). Ferroptosis was the only Enrichr derived pathway that was associated with non ACE2 vector samples due to context-specific essential genes (SLC3A2, SLC7A11). Protein digestion and absorption were also enriched in some conditions with high viral load. Whereas, glycerophospholipid metabolism was highly enriched in both lung study. Finally, Pyrimidine metabolism was enriched in most severity study conditions, meanwhile, sphingolipid metabolism was enriched in all time-series study conditions.

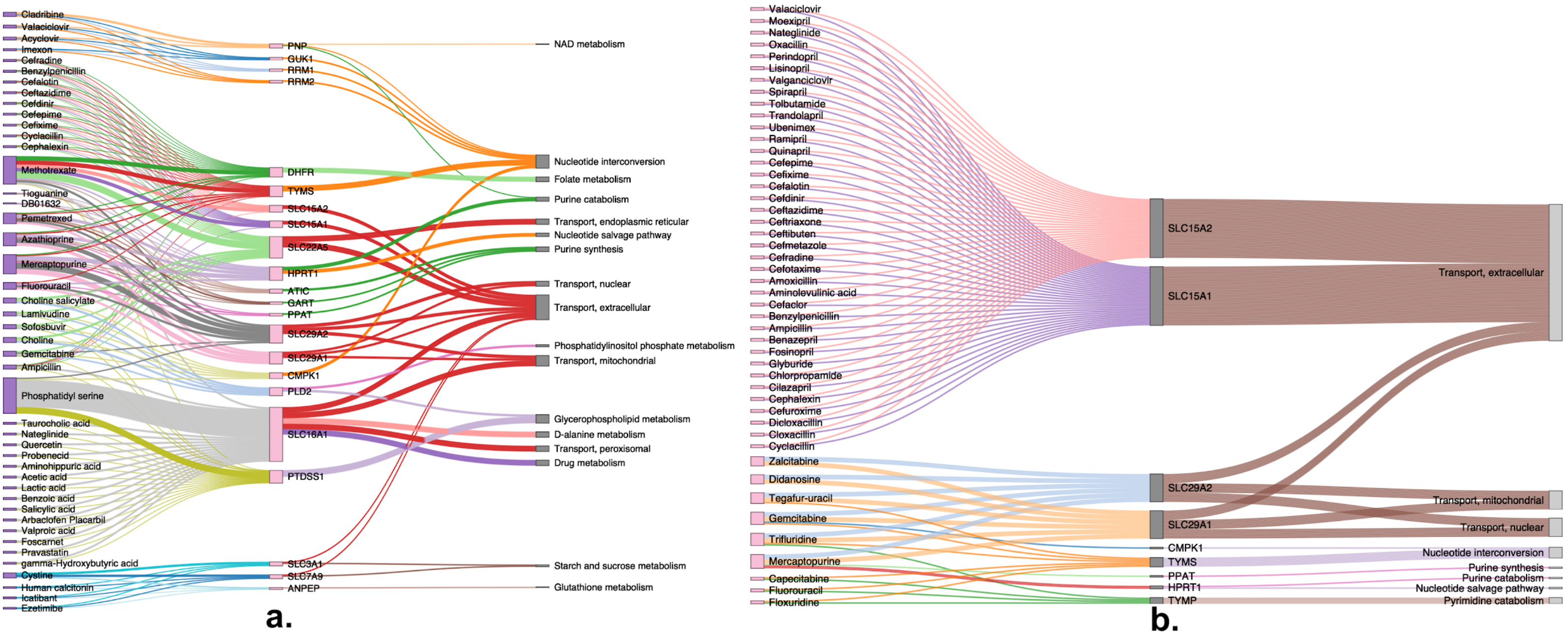

**Fig. 5: Tripartite Network of the Drug-Gene-Pathway Interactions of the Synergetic Gene-pairs Determined by Double Gene Deletion.**
Determination for the individual repositioned drugs for the synergistic gene-pairs, and also mapping the relationships of the genes to pathways determined by Recon3D_01 subSystem. a) Tripartite network of the reduced list of Double Gene Deletion Drug Pairs: Candidate gene pairs with possible essentiality to the viral biomass were determined by double gene knockout (DKO) as mentioned in (A.2.3). Gene pairs that have one drug for both genes were excluded. Genes, and their connected drugs, that don't have pathways in Recon3D_01 subSystems were discarded. b) Tripartite network of Single Drugs with Two Paired Targets of Double Gene Deletion: Candidate genes with possible essentiality to the viral biomass were determined by double gene knockout as mentioned in (A.2.3). Gene pairs that have one drug for both gene targets were selected. Genes, and their connected drugs, that don't have pathways in Recon3D_01 subSystems were discarded.

## Prediction of candidates for repositioning of drugs and drug combinations targeting essential genes and synergetic gene-pairs

Out of the 23 predicted essential genes, eight genes are druggable by 45 unique drugs (**Table S3**) from DrugBank [58]. Six antiviral drugs (acyclovir, valaciclovir, lamivudine, sofosbuvir, methotrexate, trifluridine) were identified in these 45 drugs. These drugs cover many pharmacological actions such as immunosuppressive, antiviral, folic acid antagonists, antirheumatic, and hypolipidemic actions besides some known nutraceutical cofactors such as lactose and folic acid. The tripartite network of individual repositioned drugs (Fig. 5 b) shows a multi-target effect of four drugs (pralatrexate, pemetrexed, methotrexate, Gemcitabine). Gemcitabine affects the nucleotide interconversion pathway through both CPMK1 and TYMS essential genes. Meanwhile, pralatrexate, pemetrexed and methotrexate, affect both nucleotide interconversion and folate metabolism pathways through TYMS and DHFR essential genes respectively.

Double gene deletion produced 598 unique gene pairs across the two lung studies. Out of these 598 gene-pairs, 56 gene-pairs are druggable by 3411 unique drugs or drug pairs. We found 47 single drugs with two paired targets, (**Table S4**) due to multiple identified targets per drug. Since these 3411 drug pairs could target more than one gene pair, safety scores and essentiality scores were calculated using the average of these scores. To prioritize among the 3411 drug pairs, we filtered by keeping drug pairs with more than two in essentiality scores, and more than one in either the number of gene-pairs or safety scores. This reduced drug pair list has 52 drug pairs that consist of 37 individual drugs (**Table S5**). The top-ranked drug pairs in the number of gene pairs are (azathioprine-pemetrexed and mercaptopurine-pemetrexed) affecting five essential gene pairs, while (imexon-valaciclovir and imexon-acyclovir) are top-ranked in the essentiality score of 10. The pathway analysis of the druggable essential gene-pairs (see **Fig. 5**), shows that most of the single drugs with two paired targets, targets the extracellular transport pathway. Meanwhile the reduced drug pairs cover more diverse pathways. These pathways include new pathways in addition to the single druggable pathways such as: purine catabolism, purine synthesis, nucleotide salvage pathway, and NAD metabolism. The aforementioned azathioprine-pemetrexed drug pair targets seven metabolic pathways such as transporter pathway and purine synthesis and catabolism.

## Discussion

In the present study, we analyzed changes in transcriptomic data of lung cell lines infected with COVID-19 at various viral loads and at different time points after infection. The main focus was on the alteration of expression of metabolic genes that could be evidence of a metabolic rewiring induced by the virus. Then *in silico* single and gene double knockouts were performed to identify potential infected cell-specific essential genes that arise from this metabolic rewiring and that could be used as potential drug targets. To extend the list of targets and identify critical pathways for the growth or survival of the virus, reactions were inactivated *in silico* and the resulting impact on the viral biomass production was estimated. In addition, we explored pathways enriched for predicted essential genes and reactions to obtain a better picture of the occurring metabolic rewiring. Furthermore, we predicted a set of 45 repositionable drugs and 99 drug combinations that could be explored as treatment against COVID-19. Finally, we compared our results against two recent studies that cover the candidate metabolic pathways alternation in COVID-19 infection [62] [63].

To unravel the metabolic rewiring induced by COVID-19 on the host lung cell after ACE2 or a mock transfection at different viral loads, we computed the metabolic differentially expressed genes (DEGs) in the severity study for these different conditions. The DEGs were classified into three sets: downregulated with ACE2 vector, downregulated with high viral load, and moderately downregulated across all conditions.

Among the metabolic pathways with a different activation pattern after the transfection with the ACE2 vector, are chondroitin sulfate degradation and synthesis that have a link to oxidative stress. Chondroitin is a glycosaminoglycan (GAG) that plays a role as a structural component in the cartilage. It also has antioxidant and neuroprotective effects against oxidative stress through the upregulation of phosphoinositide 3-kinases (PI3K) /Akt signaling and heme oxygenase-1 (HMOX1) [64]. Chondroitin sulfate degradation was downregulated after the transfection with ACE2 vector and chondroitin synthesis was downregulated with a high viral load which is consistent with the hypothesis of accumulation chondroitin in the infected cells to balance the oxidative stress induced by the virus. The phosphoinositol phosphate pathway, which includes PI3K was also downregulated after transfection with the ACE2 vector, further supporting the protective role of chondroitin in COVID-19 infection. Although this hypothesis was not tested *in vitro,* in an *in vivo* study on the Vero cell line, chondroitin sulfate showed a weak inhibition of SARS-CoV-2 cell entry in comparison to other types of GAGs such as heparin and enoxaparin [65]. Finally, HMOX1 has been found to bind to SARS-CoV-2 open reading frame 3 a (ORF3a) [66]. The ineffectiveness of chondroitin sulfate as an antiviral agent SARS-CoV2 was expected due to its unspecific mode of action. Reducing oxidative stress may alleviate the symptoms but does not kill the virus nor does it reduce its ability to replicate itself.

To be more effective, drug candidates have to target genes, reactions, or pathways that are key and specific to viral metabolism. Hence *in silico* single and double gene knockouts were performed to identify genes essential to the viral biomass production but whose knockout has little or no effect on the host biomass production. Among the 23 predicted essential genes for viral biomass, two genes of the phospholipid metabolism (CRLS1 and SGMS1) showed the highest essentiality score. GUK1, which was the main essential gene identified in [47], is also among the top predicted targets and displays moderate essentiality and safety score. Furthermore, three essential genes (DTYMK, CMPK1 & TYMS) are part of pyrimidine metabolism (through pyrimidine deoxyribonucleotides *de novo* biosynthesis) that was enriched in all conditions in the severity study, but for the last time point of the time-series study in the Calu-3 cell line. In two separate *in vivo* studies on the Vero cell line, inhibition of *de novo* pyrimidine biosynthesis pathway through dihydroorotate dehydrogenase (DHODH) inhibitors, showed broad-spectrum antiviral activity, stopped or halted SARS-CoV-2 replication respectively [67], [68]. Although DHODH inhibitor PTC299 showed a little cytotoxic effect on SARS-CoV2, they were proven to have an immunomodulating effect on IL-6, IL-7A, IL-17F, and VEGF [68]. Alterations in the expression of genes of the pyrimidine metabolism were significantly higher in the A549, suggesting a response that could be specific to this cell line.

To further understand why a gene is essential for the viral biomass production, we examined the pathways harbouring the essential genes. A percentage of shared essential metabolic pathways were found between the two lung studies (8 out of 13 pathways), highlighting the robustness of the predictions (fatty acid oxidation, glycerophospholipid metabolism, inositol phosphate metabolism, methionine and cysteine metabolism, nucleotide interconversion, sphingolipid metabolism, starch and sucrose metabolism, extracellular transport). Pathways harbouring essential genes were also enriched for DEGs. Exploring the ferroptosis specific database FerrDb shows that three out of the 23 essential genes are related to ferroptosis. SLC7A11 and SLC3A2 are classified as suppressors, meanwhile, SLC7A5 is a marker. Also, FerrDb, classified only one of our predicted drugs, sulfasalazine, as an inducer of ferroptosis.

Ferroptosis is an iron-dependent programmed cell death that can be inhibited by Selenium. In a cross-sectional study, Selenium level was found higher in tissue samples from COVID-19 survivors in comparison to non-survivors [69]. Also in a population retrospective analysis, the selenium concentration in the hair in the population of Chinese cities outside Hubei province was correlated with the COVID-19 cure rate in Chinese cities [70]. Even though the last retrospective study using city population-level data instead of patient-level data might be less reliable, both suggest a role of ferroptosis in the survival of COVID patients.

To prioritize drug and drug combinations and since many conditions in the time-series study were lacking mock samples, we relied for the present work rather on the essentiality score of each gene identified in terms of reducing the viral proliferation rather than the predicted toxicity on control tissue models (safety score). In total 23 SARS-COV-2-specific essential genes were predicted by rFASTCORMICS based lung models, that can be targeted by 45 repositionable drugs and 99 drug combinations. The safety of the drugs was assessed by simulation knockouts on the biomass of the counterpart mock sample. This strategy allows estimating which drugs might be potential candidates as not having too drastic side-effects. Although the drug candidates are all FDA-approved drugs, some treatments are associated with severe side effects, and combining two drugs can have additional unexpected side effects. Hence, further tests would be required on other tissues and using other optimization functions as well as *in vitro* and *in vivo* validations before considering any predictions as potential drug candidates.

Among the predicted drugs, six are antiviral drugs and three drugs (valaciclovir, methotrexate, sofosbuvir) are broad-spectrum antivirals [71]. In a small clinical trial (n= 62), the combination of sofosbuvir-daclatasvir decreased the COVID-19 mortality rate (6%) in comparison to ribavirin (33%) [72]. Of the predicted drugs, two drugs (acyclovir, valaciclovir) target the GUK1 gene, which shows relative essentiality and safety. SGMS1, one of our top drug target candidates, has currently no associated drug in DrugBank. Acetylcysteine, another predicted drug by our workflow, is mucolytic and antioxidant in high doses through regenerating glutathione. Acetylcysteine has been proposed as a treatment for COVID-19 [73] for its role in inflammation regulation. Acetylcysteine alone or with bromelain were able *in vitro* to fragment the recombinant spike and envelope SARS-CoV-2 proteins [74]. Also, acetylcysteine was among four metabolic cofactors that reduced the recovery time in combination with hydroxychloroquine in comparison to hydroxychloroquine alone in a randomized clinical study (n=93) [75].

Further, statins, lipid-lowering drugs that were enriched among the predicted drugs were debated for their efficacy in reducing COVID-19 severity at the onset of the pandemic and their usefulness for COVID-19 is still unclear [76]. For example, statins were described to increase ACE2 expression through angiotensin 2 [77], which might enhance viral cell entry. On the other hand, it can decrease oxidative stress from ACE-2 viral binding [76]. Also, a retrospective study (n=13,981) has shown an association between statins and reduced COVID-19 mortality from 9.4% in patients not taking statins to 5.2% with statins [78]. Due to the relative number of the different statin recipients, this study couldn't rank the different statin types. But, a recent in-vitro study of different statins showed an antiviral effect on SARS-CoV-2 [79]. Fluvastatin was the only statistically significant in its antiviral effect, meanwhile, rosuvastatin was ranked second in the antiviral activity [79]. Moreover, gemcitabine has been shown to have antiviral activity against SARS-CoV-2 in the Vero-E6 cell line [80]. Till 17 Sep 2020, out of the 45 essential genes' drugs, seven drugs are being tested currently in clinical trials (rosuvastatin, acetylcysteine, sofosbuvir, liothyronine, melphalan, oleic acid, methotrexate) according to DrugBank COVID-19 Clinical Trial Summary [58] [81].

Also, among the 47 predicted single drugs with two paired targets, four drugs are affecting more than one gene pair (gemcitabine, trifluridine, mercaptopurine, tegafur-uracil). In the reduced list of predicted drug combinations, the highest ranked drug pair in the number of target gene pairs are azathioprine-pemetrexed and mercaptopurine-pemetrexed, both affecting five gene pairs. Furthermore, nine predicted drugs belonging to angiotensin-converting enzyme inhibitors (ACEIs) such as lisinopril, are affecting the gene pair SLC15A1-SLC15A2, targeting the extracellular transport pathway. Interestingly, in a prospective study of COVID-19 (n=19,486), patients taking ACEIs have a reduced risk of COVID-19, with differences according to ethnicity [82]. Meanwhile, ACEIs did not reduce the risk of receiving ICU care [82].

In both drug pair datasets (single drugs with two paired targets, see reduced list), immunomodulators appear such as mercaptopurine, azathioprine, pemetrexed, and methotrexate. Methotrexate shows antiviral activity against SARS-CoV-2 in Vero-E and Calu-3 cell lines [83]. This antiviral activity was better than the only authorized antiviral for emergency use for COVID-19 remedisvir. Also, five unique antivirals appear in the two drug pairs datasets (didanosine, trifluridine, valganciclovir, valaciclovir, zalcitabine).

The use of 5-aminosalicylate or sulfasalazine, a 5-aminosalicylate prodrug, has recently been shown to increase COVID-19 severity in patients with inflammatory bowel disease (IBD) in a retrospective study (n = 525) [84]. As sulfasalazine was found as a ferroptosis inducer in FerrDb [85], this could strengthen the evidence of a role of ferroptosis in COVID-19 severity. It illustrates also the need for a careful assessment of the toxicity of the predicted drugs in follow-up *in vitro* and *in vivo* studies as well as a patient or group of patient-tailored approach as some of the most efficient treatments for a given group of patients is likely to have severe adverse side-effects for others.

We further compared our enriched pathways from the DEGs and essential genes with a recent metabolomics study [62]. In this study serum metabolites were compared among SARS-CoV-2 positive and negative patients, also altered IL-6 levels measures as an indication of COVID-19 severity. The study found several altered pathways and dysregulation notably of nitrogen and tryptophan metabolism associated with increased severity. Also, some metabolite levels were increased in COVID-19 patients such as kynurenines, methionine sulfoxide, cystine, and free polyunsaturated fatty acids (PUFAs).

Pathways that harbored essential genes (essential gene pathways) and DEGs (**Fig. 2**) between COVID-19 and mock samples were also identified as relevant pathways in the metabolomics study [62]. Up-regulated fatty acid oxidation in DEGs and glycerophospholipid metabolism in essential genes pathways is consistent with the increased levels of PUFAs [62]. The increased PUFAs are biomarkers for ferroptosis which was predicted in the condition without ACE2 vector. The recent evidence for the role of selenium in COVID-19 and the significant presence of PUFAs as a biomarker in severe COVID-19 cases might be a further indication of the role of ferroptosis in regards to COVID-19 severity. Moreover, the SARS-CoV-2 spike protein was discovered to have a binding pocket for free fatty acids [86]. This seems to allow the PUFA linoleic acid to have a synergistic effect with the antiviral remdesivir against SARS-CoV-2 *in vitro* [86].

Furthermore, tryptophan metabolism that was up-regulated in the DEGs in conditions with high viral load (A549_2, A549_2_ACE2), has been found to be dysregulated in the metabolomics study. Also, the enriched methionine and cysteine metabolism identified in (**Fig. 4**) might explain the increased oxidative stress biomarker levels of methionine sulfoxide and cystine and the enriched nitrogen metabolism in the metabolomics study [62].

To discover which lung essential pathways might be shared with other infected organs, we also compared our identified pathways with a multi-omics study on three cell types: megakaryocytes, erythroid cells, and plasmablasts [63]. In this longitudinal study of COVID-19 severity, cell-/tissue- specific metabolic models were reconstructed from single-cell/ bulk RNA-seq respectively [63]. The goal of the metabolic reconstruction in this study was to find cell-specific metabolic pathways associated with different disease progression and recovery time points [63]. The essentiality of genes and reactions in these pathways across the three cell types are unknown since single gene or reaction deletions were not applied. Many identified metabolic pathways in this multi-omics study across megakaryocytes, erythroid cells, and plasmablasts were also shared with our essential pathways on the lung such as pyrimidine metabolism and cysteine and methionine metabolism. Also, our lung essential metabolic pathways such as inositol phosphate metabolism, sphingolipid metabolism, glycine, serine, and threonine metabolism have been identified as both erythroid cells- and plasmablasts- specific. Meanwhile, the lung- essential fatty acid oxidation and the non-essential pyruvate metabolism have been identified as megakaryocytes- specific. Interestingly, a high upregulation of pyruvate kinase M in (PI3K) /Akt signaling was found in critical patients in megakaryocytes [63]. (PI3K) /Akt signaling participates in chondroitin sulfate metabolism [64], which was enriched in the DEGs but not in the essential pathways. Furthermore, serum sphingosine-1-phosphate, a metabolite in the lung- essential sphingolipid metabolism, was found significantly decreased with COVID-19 severity in a small study (n=111) [87]. Taken together, the shared metabolic pathways between the different studies such as pyrimidine metabolism and methionine and cysteine metabolism across different tissues might represent core viral-specific pathways that could harbor efficient drug targets that would eliminate or slow down the virus regardless of the infected tissue.

Although this study predicts some interesting drug candidates and drug combinations, the work is limited by the modelled lung cell lines (A549, Calu-3, H1299, NHBE). Since more evidence shows the multi-organ effect of COVID-19 [40], especially the possible long-term effects [44], further computational studies are needed e.g., to obtain multicellular constraint-based models [88]. With the rising SARS-CoV-2 variants globally, variant-specific metabolic models would be needed to answer questions related to vaccines' immunity-escaping. Another limitation to this work is that the identified drug and drug pairs are based on targets identified by network effects on the host metabolome (as the virus is only modelled through its biomass function) rather than direct docking on the viral proteome. Thus, further *in vitro* studies are needed to determine the essentiality of the identified essential genes.

## Conclusion

Unlike drug repositioning using expression reversal or drug docking that lack targets' identification or genome-scale multi-targeting respectively, constraint-based metabolic modeling is a powerful *in silico* tool for drug repositioning with genome-scale information and producing known targets. These powerful advantages come from gene essentiality prediction. In this work, context- specific models from expression data from infected lung cell lines were built then constrained by both viral and host biomass. *In silico* gene deletion identified 23 single essential genes and 598 essential gene pairs. Drug repositioning using approved drugs in DrugBank identified 45 single drugs, and 99 drug combinations of which 47 single drugs are targeting both genes in the gene pair. Pathway analysis of the essential genes identify ferroptosis as a candidate biomarker pathway of COVID-19 severity. Gemcitabine was predicted to target two single essential genes in the nucleotide interconversion pathway and three gene pairs in drugs identified by both single and double gene deletion respectively. Finally, we predicted the GUK1 gene as both relatively safe and essential against SARS-CoV2 as reported by a previous *in silico* modeling.

### Data Availability Statement

Models and code are available in:

https://github.com/sysbiolux/DCcov

FPKM for the time-series study GSE148729:
https://filetransfer.mdc-berlin.de/?u=CVXckugR&p=MACT6Xw9

GHDDI Broad-spectrum antiviral agents:
https://ghddiai.oss-cn-zhangjiakou.aliyuncs.com/file/Antivirus_Drug_Profile_k2.csv

**Acknowledgments:**

The experiments presented in this paper were carried out using the HPC facilities of the University of Luxembourg [89].

## Supplementary Information:

re

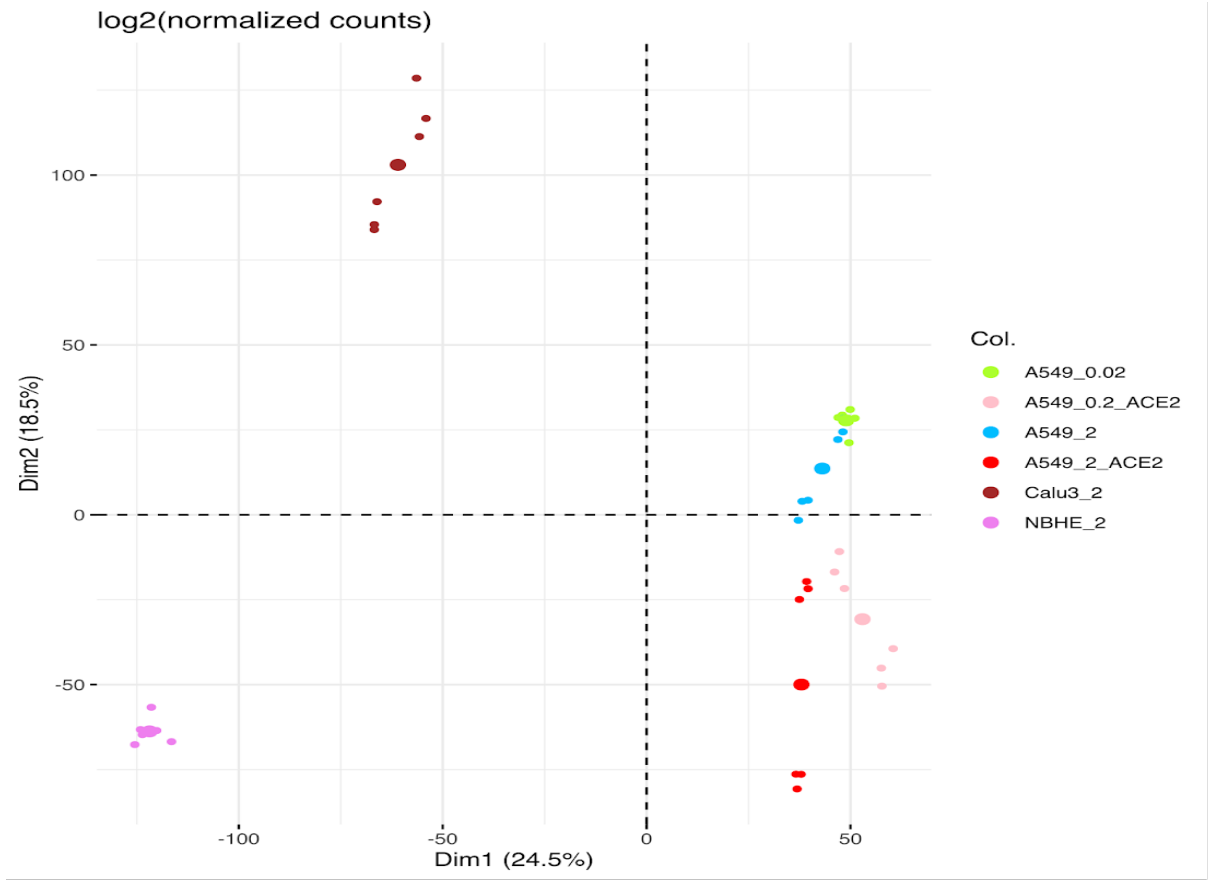

a

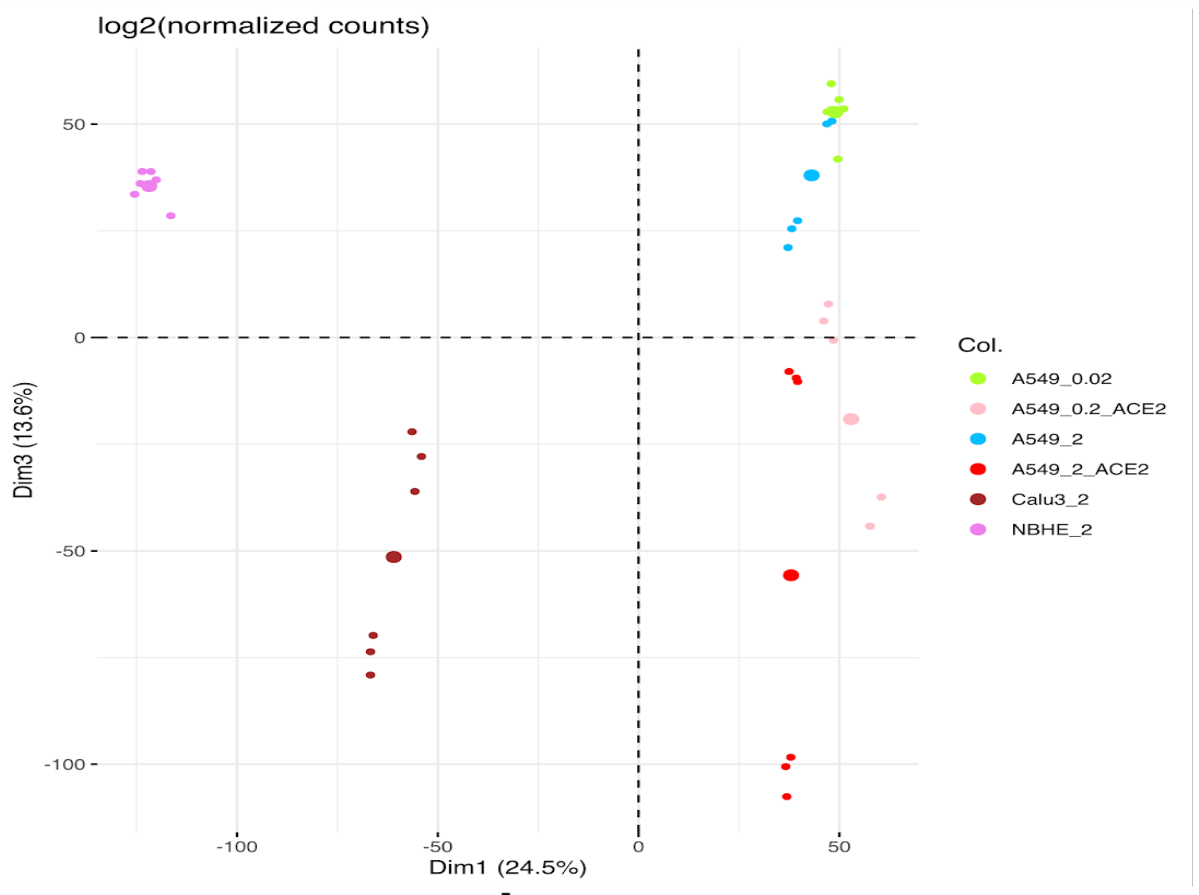

b

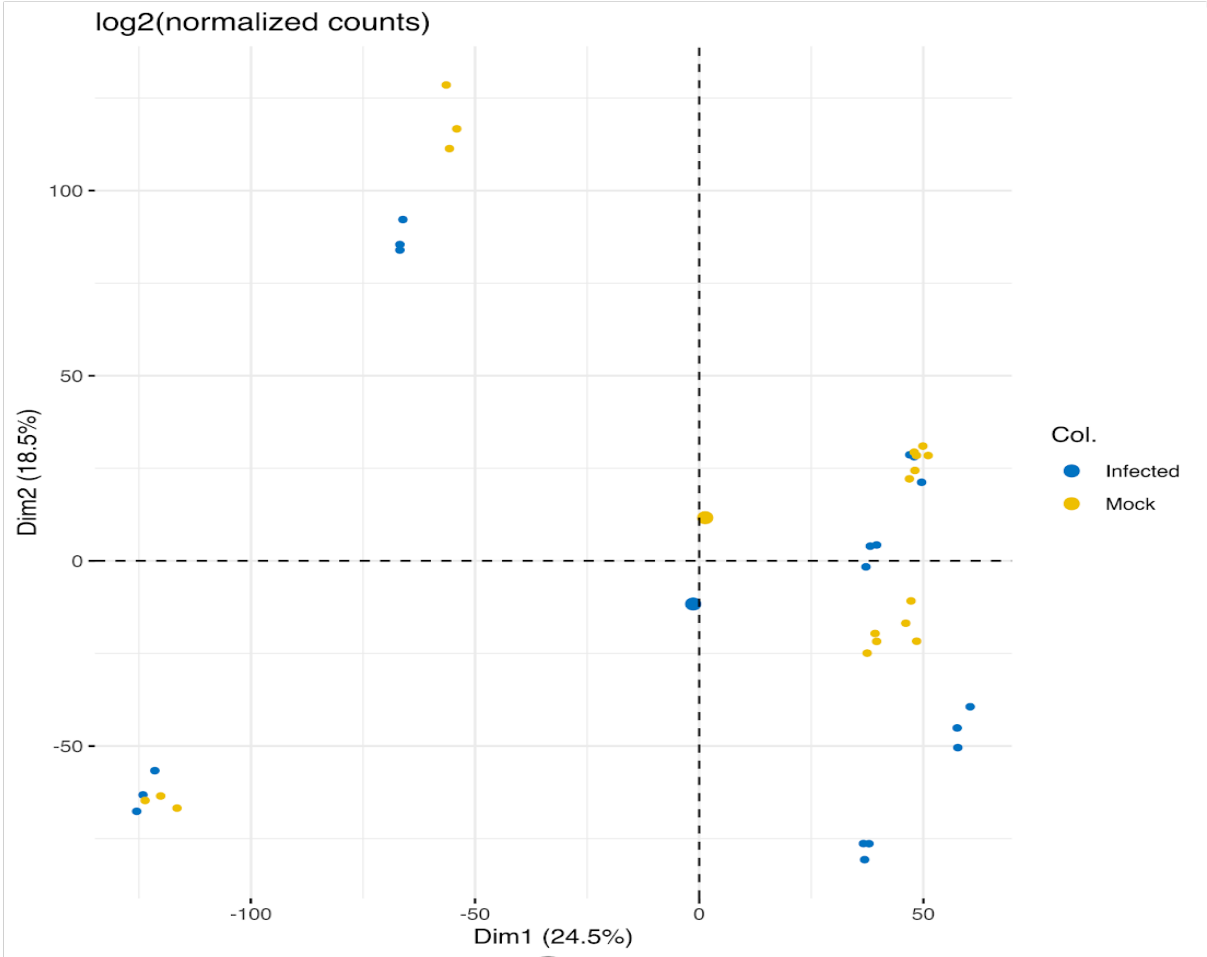

c

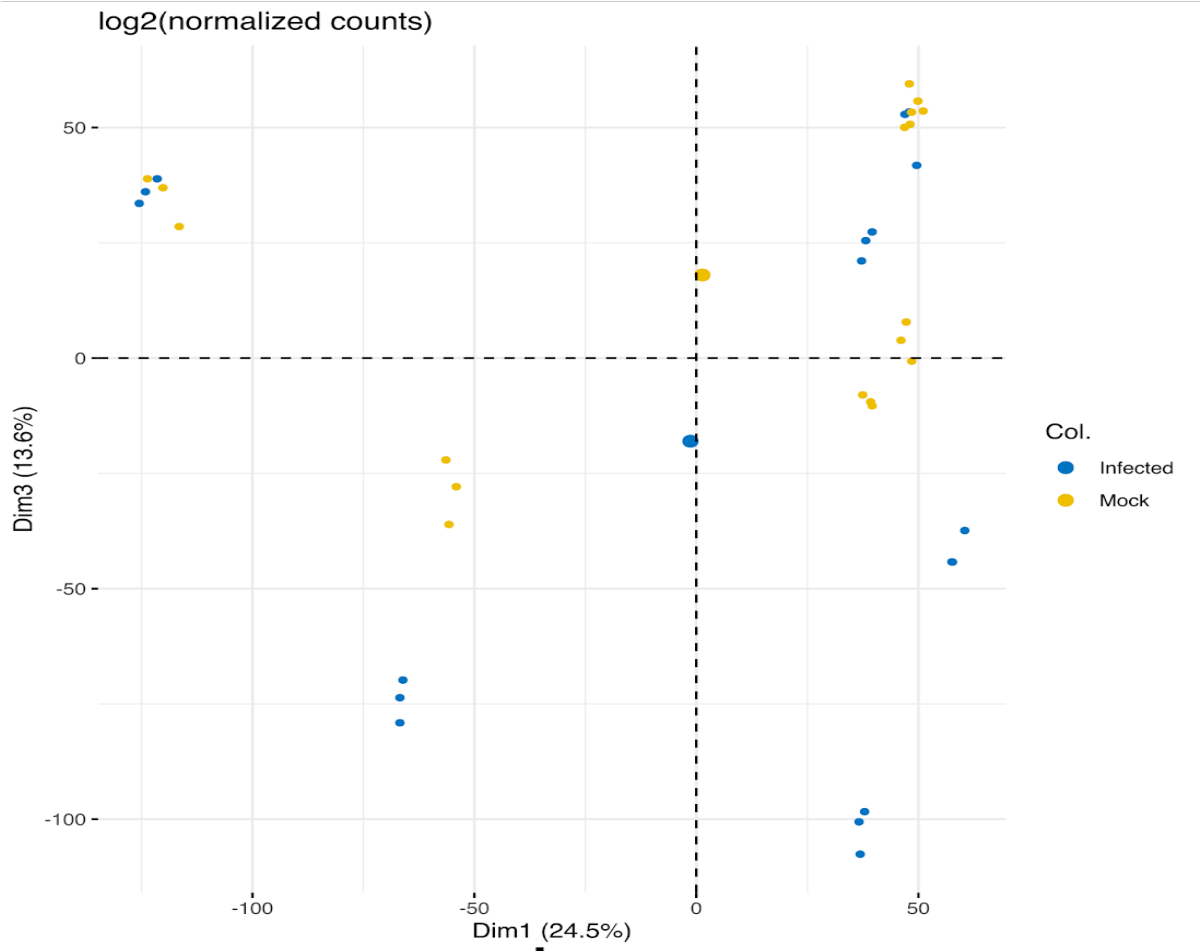

d

**Fig. S1: Principal Component Analysis (PCA) of the Severity Study given in Log Normalized Read Counts for different principal components (PC).**

a) PC1 vs PC2 colored by cell line and condition.
b) PC1 vs PC3 colored by cell line and condition.
c) PC1 vs PC2 colored by infection status (infected vs. mock samples).
d) PC1 vs PC3 colored by infection status (infected vs mock samples).

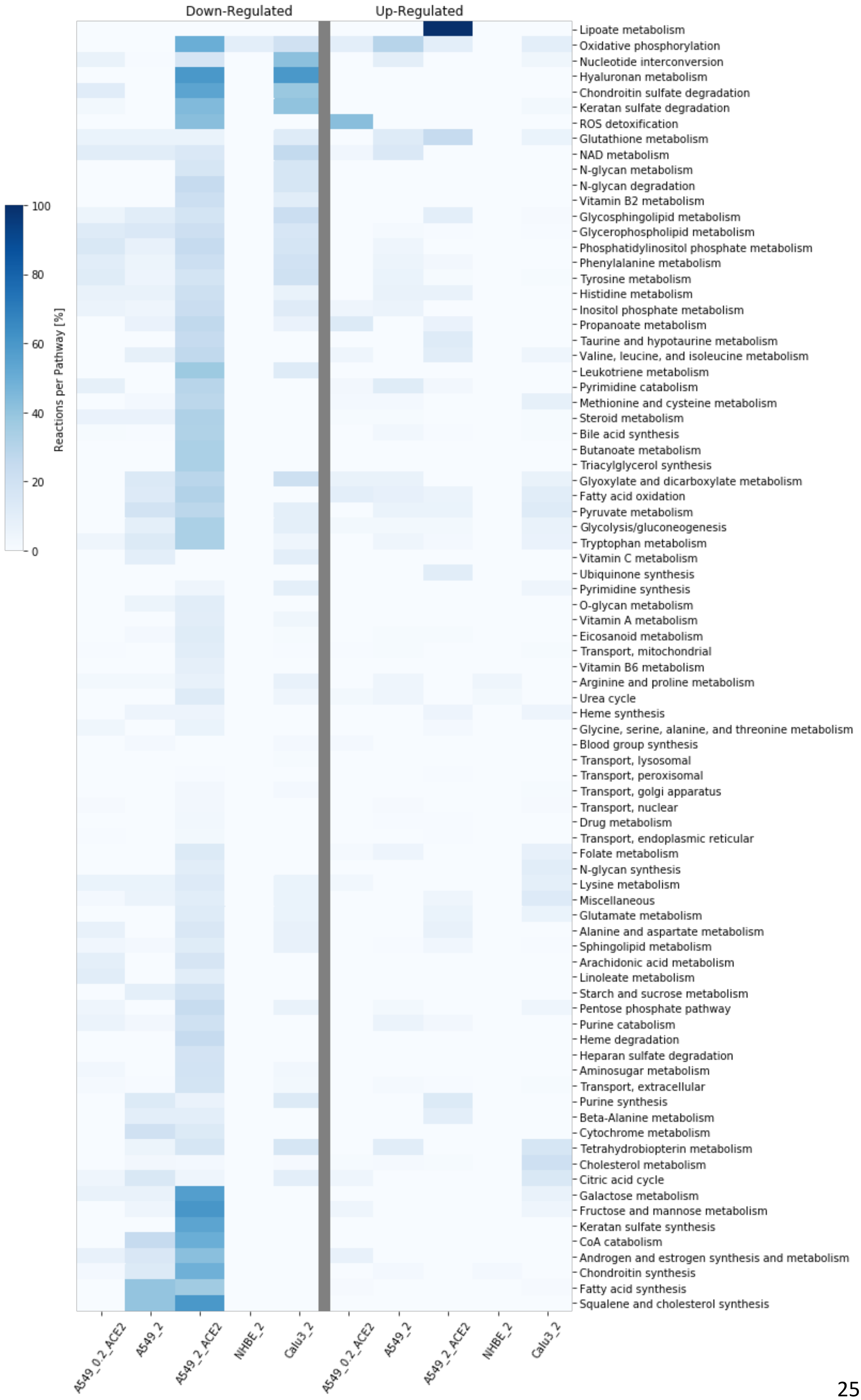

**Fig. S2**: **Reactions per Pathway Heatmap for Pathway Analysis of All Conditions of Differentially expressed genes in the Severity Study.** Differentially expressed genes (DEGs) were computed with DESeq2. The Down- and up-regulated were mapped to the pathways (subSystems) of Recon3D_01. The number of up and down-regulated reactions was then summed up to identify the top altered pathways in the infected lung cell lines in the severity study (A.1.3) but without filters on the number of reactions nor the Reactions per Pathway. The color code "Reactions per Pathway [%]" represents the number of enriched metabolic reactions in a pathway divided by the overall number of reactions in this pathway. Removing filters allowed increased the number of differentially expressed pathways and allowed including the NBHE_2 condition.

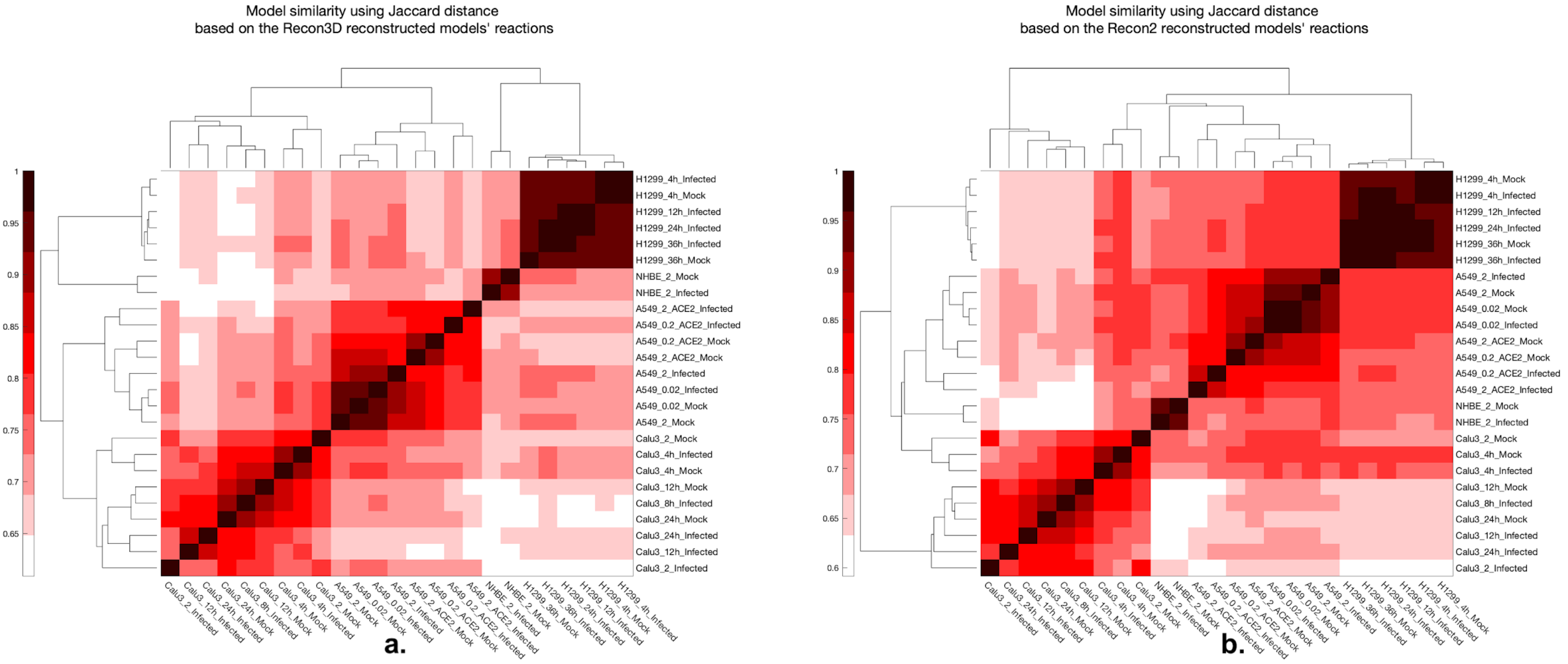

**Fig. S3: Clustergrams of the Reconstructed Models' Similarity using Reactions Presence with Jaccard Similarity Metrics of Both Recon3D Models (a) and Recon2 Models (b).**

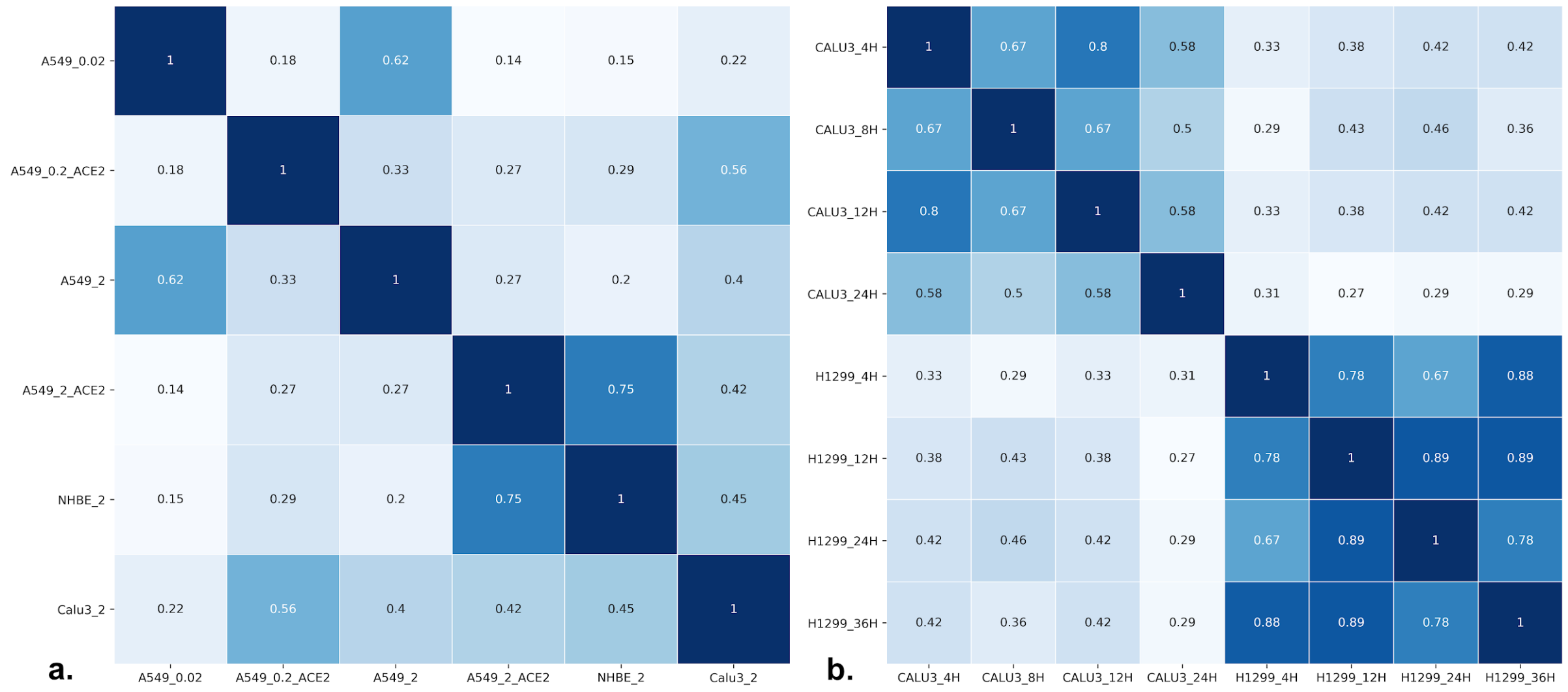

**Fig. S4: Jaccard Similarity of the Essential Genes derived from the different conditions of the 2 lung studies (Severity Study & Time-series Study)**

**Table S1: Essential metabolic genes of the severity study.** The essential genes of each condition of the infected model are aggregated from the Recon2.04 and Recon3D_01 based reconstructions. Genes are classified according to the predicted toxicity into toxic

or safe, based on the presence or the absence from the essential genes of the mock model. A gene is considered with unknown toxicity, if there is no mock model corresponding to the infected model, or the essential gene was filtered from the context mock model.

| Condition | Safe Essential | Toxic Essential | Unknown Safety Essential |
|---|---|---|---|
| **A549_0.02** | SLC3A2; SLC7A11 | CRLS1; ISYNA1; SGMS1 | |
| **A549_0.2_ACE2** | AGXT; DHFR; SLC27A4; TYMS | CRLS1; GUK1; PEPD; SGMS1 | |
| **NHBE_2** | SLC3A1; SLC7A9 | CMPK1; CRLS1; CTH; GUK1; PTDSS1; SGMS1; SLC27A4; SLC5A3 | SLC5A3 |
| **A549_2** | DTYMK; GUK1; PEPD | CRLS1; GUK1; ISYNA1; SGMS1; SLC3A2; SLC7A11 | |
| **A549_2_ACE2** | CMPK1; CTH; GUK1; PTDSS1; SLC3A1; SLC7A9 | CRLS1; PEPD; SGMS1; SLC2A13 | SLC5A3 |
| **Calu3_2** | CMPK1; SLC27A4 | CRLS1; GUK1; PEPD; SGMS1 | |

**Table S2: Essential metabolic genes of the time-series study**
The essential genes of each condition of the infected model are aggregated from the Recon2.04 and Recon3 based reconstructions. Gene gain and gene loss are genes gained or lost from the previous time point in the same cell line. Time points are in hours. Genes are classified according to the predicted toxicity into toxic or safe, based on the presence or the absence from the essential genes of the mock model. A gene is considered with unknown toxicity, if there is no mock model corresponding to the infected model, or the essential gene was filtered from the context mock model.

| Condition | Safe Essential | Toxic Essential | Unknown Safety Essential | Gene Gain | Gene Loss |
|---|---|---|---|---|---|
| **Calu3_4H** | GLTP | CRLS1; GUK1; ISYNA1; PEPD; SGMS1; SLC27A4; SLC3A2; SLC7A11 | | | |
| **Calu3_8H** | | | CRLS1; GLTP; GUK1; ISYNA1; PEPD; SGMS1; SLC27A4; SLC3A2; SLC7A11; SLC7A5 | SLC7A5 | |
| **Calu3_12H** | ISYNA1; SLC27A4; SLC7A5 | CRLS1; GUK1; ISYNA1; PEPD; SGMS1; SLC3A2; SLC7A11; SLC7A5 | | | GLTP |
| **Calu3_24H** | PTDSS1 | CRLS1; GLTP; GUK1; ISYNA1; SGMS1; SLC27A4; SLC3A2; SLC7A11; SLC7A5 | | GLTP; PTDSS1 | |
| **H1299_4H** | CTH | CRLS1; ISYNA1; PLD2; SGMS1; SLC27A4; SLC7A6 | | | |
| **H1299_12H** | | | CRLS1; CTH; ISYNA1; PEPD; PLD2; SGMS1; SLC27A4; SLC7A6 | PEPD | |
| **H1299_24H** | | | CRLS1; ISYNA1; PEPD; PLD2; SGMS1; SLC27A4; SLC7A6 | | CTH |
| **H1299_36H** | PEPD; SLC7A6 | CRLS1; CTH; ISYNA1; PLD2; SGMS1; SLC27A4 | | CTH | |

**Table S3: Candidate drugs for the essential genes given by their gene targets.** Drug-target interactions were downloaded from DrugBank and filtered as mentioned in (C). Essential genes were determined using single-gene knockouts. eight essential genes are targeted by 45 drugs.

| Gene Symbol | Drugs | Essentiality Score | Safety Score |
|---|---|---|---|
| **GUK1** | Valaciclovir, Acyclovir | 3 | 16 |
| **ISYNA1** | Glycerin | 1 | 9 |
| **SLC7A11** | Riluzole, Taurocholic acid, Alanosine, Thimerosal, Tauroursodeoxycholic acid, Rosuvastatin, Acetylcysteine, Sulfasalazine | 1 | 5 |
| **CMPK1** | Lamivudine, Gemcitabine, Sofosbuvir | 3 | 4 |
| **SLC7A5** | Dextrothyroxine, Pregabalin, Levothyroxine, Liothyronine, Melphalan, Gabapentin, Levodopa, Thyroid, porcine | 1 | 3 |
| **GLTP** | Oleic Acid, Lactose, Lauric acid | 1 | 2 |
| **DHFR** | Methotrexate, Pyrimethamine, Pemetrexed, Lamotrigine, Piritrexim, Pralatrexate, Trimetrexate, Proguanil, Iclaprim, Gentamicin | 1 | 1 |
| **TYMS** | ANX-510, Fluorouracil, Capecitabine, Methotrexate, Floxuridine, Thymectacin, Trifluridine, Gemcitabine, Pemetrexed, Pralatrexate, OSI-7904L, Tegafur, Tegafur-uracil, Raltitrexed | 1 | 1 |

**Table S4: Single Drugs with Two Paired Targets of Double Gene Deletion:** Candidate gene pairs with possible essentiality to the viral biomass were determined by double gene knockout as mentioned in (A.2.3). Gene pairs that have one drug for both gene targets were selected. Essentiality and safety scores were calculated as mentioned in A.2.4. As many gene-pairs share the same drug, gene pairs were clustered by their drugs. Essentiality and safety scores were averaged between the different gene-pairs. Drugs were sorted by the number of gene pairs they are targeting.

| Drug | Gene Pairs | Average Essentiality Score | Average Safety Score | Number of Gene Pairs |
|---|---|---|---|---|
| Gemcitabine | CMPK1;SLC29A2, TYMS;SLC29A2, SLC29A2;SLC29A1 | 1.66 | 1.0 | 3 |
| Trifluridine | TYMS;TYMP, TYMS;SLC29A2, SLC29A2;SLC29A1 | 1.33 | 1.0 | 3 |
| Mercaptopurine | PPAT;HPRT1, SLC29A2;SLC29A1 | 3.5 | 1.0 | 2 |
| Tegafur-uracil | TYMS;SLC29A2, SLC29A2;SLC29A1 | 1.5 | 1.0 | 2 |
| Valaciclovir | SLC15A2;SLC15A1 | 13.0 | 0.0 | 1 |
| Moexipril | SLC15A2;SLC15A1 | 13.0 | 0.0 | 1 |
| Nateglinide | SLC15A2;SLC15A1 | 13.0 | 0.0 | 1 |
| Oxacillin | SLC15A2;SLC15A1 | 13.0 | 0.0 | 1 |
| Perindopril | SLC15A2;SLC15A1 | 13.0 | 0.0 | 1 |
| Lisinopril | SLC15A2;SLC15A1 | 13.0 | 0.0 | 1 |
| Valganciclovir | SLC15A2;SLC15A1 | 13.0 | 0.0 | 1 |
| Spirapril | SLC15A2;SLC15A1 | 13.0 | 0.0 | 1 |
| Tolbutamide | SLC15A2;SLC15A1 | 13.0 | 0.0 | 1 |
| Trandolapril | SLC15A2;SLC15A1 | 13.0 | 0.0 | 1 |
| Ubenimex | SLC15A2;SLC15A1 | 13.0 | 0.0 | 1 |
| Ramipril | SLC15A2;SLC15A1 | 13.0 | 0.0 | 1 |
| Quinapril | SLC15A2;SLC15A1 | 13.0 | 0.0 | 1 |
| Cefepime | SLC15A2;SLC15A1 | 13.0 | 0.0 | 1 |
| Cefixime | SLC15A2;SLC15A1 | 13.0 | 0.0 | 1 |
| Cefalotin | SLC15A2;SLC15A1 | 13.0 | 0.0 | 1 |
| Cefdinir | SLC15A2;SLC15A1 | 13.0 | 0.0 | 1 |
| Ceftazidime | SLC15A2;SLC15A1 | 13.0 | 0.0 | 1 |
| Ceftriaxone | SLC15A2;SLC15A1 | 13.0 | 0.0 | 1 |
| Ceftibuten | SLC15A2;SLC15A1 | 13.0 | 0.0 | 1 |
| Cefmetazole | SLC15A2;SLC15A1 | 13.0 | 0.0 | 1 |
| Cefradine | SLC15A2;SLC15A1 | 13.0 | 0.0 | 1 |
| Cefotaxime | SLC15A2;SLC15A1 | 13.0 | 0.0 | 1 |
| Amoxicillin | SLC15A2;SLC15A1 | 13.0 | 0.0 | 1 |
| Aminolevulinic acid | SLC15A2;SLC15A1 | 13.0 | 0.0 | 1 |
| Cefaclor | SLC15A2;SLC15A1 | 13.0 | 0.0 | 1 |
| Benzylpenicillin | SLC15A2;SLC15A1 | 13.0 | 0.0 | 1 |
| Ampicillin | SLC15A2;SLC15A1 | 13.0 | 0.0 | 1 |
| Benazepril | SLC15A2;SLC15A1 | 13.0 | 0.0 | 1 |
| Fosinopril | SLC15A2;SLC15A1 | 13.0 | 0.0 | 1 |
| Glyburide | SLC15A2;SLC15A1 | 13.0 | 0.0 | 1 |
| Chlorpropamide | SLC15A2;SLC15A1 | 13.0 | 0.0 | 1 |
| Cilazapril | SLC15A2;SLC15A1 | 13.0 | 0.0 | 1 |
| Cephalexin | SLC15A2;SLC15A1 | 13.0 | 0.0 | 1 |
| Cefuroxime | SLC15A2;SLC15A1 | 13.0 | 0.0 | 1 |
| Dicloxacillin | SLC15A2;SLC15A1 | 13.0 | 0.0 | 1 |
| Cloxacillin | SLC15A2;SLC15A1 | 13.0 | 0.0 | 1 |
| Cyclacillin | SLC15A2;SLC15A1 | 13.0 | 0.0 | 1 |
| Zalcitabine | SLC29A2;SLC29A1 | 2.0 | 2.0 | 1 |
| Didanosine | SLC29A2;SLC29A1 | 2.0 | 2.0 | 1 |
| Capecitabine | TYMS;TYMP | 1.0 | 1.0 | 1 |
| Fluorouracil | TYMS;TYMP | 1.0 | 1.0 | 1 |
| Floxuridine | TYMS;TYMP | 1.0 | 1.0 | 1 |

**Table S5: Reduced list of Double Gene Deletion Drug Pairs**
Candidate gene pairs with possible essentiality to the viral biomass were determined by double gene knockout (DKO) as mentioned in (A.2.3). Gene-pairs that have one drug for both genes were excluded. Essentiality and safety scores and the number of gene-pairs were calculated such as in (**Table S4**). The reduced DKO drug dataset was filtered by selecting drug pairs with more than two in essentiality scores, and more than one in either the number of gene-pairs or safety scores.

| Drugs1 | Drugs2 | Gene Pair | Average Essentiality Score | Average Safety Score | Number of Gene Pairs |
|---|---|---|---|---|---|
| Azathioprine | Pemetrexed | HPRT1;GART, SLC29A2;SLC29A1, TYMS;SLC29A2, DHFR;SLC29A2, ATIC;HPRT1 | 2.8 | 0.4 | 5 |
| Mercaptopurine | Pemetrexed | HPRT1;GART, SLC29A2;SLC29A1, TYMS;SLC29A2, DHFR;SLC29A2, ATIC;HPRT1 | 2.8 | 0.4 | 5 |
| Cladribine | Valaciclovir | GUK1;PNP, RRM2;GUK1, RRM1;GUK1 | 9.33 | 0.0 | 3 |
| Cladribine | Acyclovir | GUK1;PNP, RRM2;GUK1, RRM1;GUK1 | 9.33 | 0.0 | 3 |
| Cefradine | Methotrexate | SLC22A5;DHFR, SLC22A5;TYMS, SLC15A2;SLC15A1 | 5.0 | 0.66 | 3 |
| Benzylpenicillin | Methotrexate | SLC22A5;DHFR, SLC22A5;TYMS, SLC15A2;SLC15A1 | 5.0 | 0.66 | 3 |
| Cefalotin | Methotrexate | SLC22A5;DHFR, SLC22A5;TYMS, SLC15A2;SLC15A1 | 5.0 | 0.66 | 3 |
| Ampicillin | Methotrexate | SLC22A5;DHFR, SLC22A5;TYMS, SLC15A2;SLC15A1 | 5.0 | 0.66 | 3 |
| Ceftazidime | Methotrexate | SLC22A5;DHFR, SLC22A5;TYMS, SLC15A2;SLC15A1 | 5.0 | 0.66 | 3 |
| Cefdinir | Methotrexate | SLC22A5;DHFR, SLC22A5;TYMS, SLC15A2;SLC15A1 | 5.0 | 0.66 | 3 |
| Cefepime | Methotrexate | SLC22A5;DHFR, SLC22A5;TYMS, SLC15A2;SLC15A1 | 5.0 | 0.66 | 3 |
| Cefixime | Methotrexate | SLC22A5;DHFR, SLC22A5;TYMS, SLC15A2;SLC15A1 | 5.0 | 0.66 | 3 |
| Cyclacillin | Methotrexate | SLC22A5;DHFR, SLC22A5;TYMS, SLC15A2;SLC15A1 | 5.0 | 0.66 | 3 |
| Cephalexin | Methotrexate | SLC22A5;DHFR, SLC22A5;TYMS, SLC15A2;SLC15A1 | 5.0 | 0.66 | 3 |
| Azathioprine | Fluorouracil | TYMS;SLC29A2, PPAT;HPRT1, SLC29A2;SLC29A1 | 2.66 | 0.66 | 3 |
| Mercaptopurine | Fluorouracil | TYMS;SLC29A2, PPAT;HPRT1, SLC29A2;SLC29A1 | 2.66 | 0.66 | 3 |
| Methotrexate | Azathioprine | TYMS;SLC29A2, DHFR;SLC29A2, ATIC;HPRT1 | 2.33 | 0.0 | 3 |
| Methotrexate | Mercaptopurine | TYMS;SLC29A2, DHFR;SLC29A2, ATIC;HPRT1 | 2.33 | 0.0 | 3 |
| Imexon | Valaciclovir | RRM2;GUK1, RRM1;GUK1 | 10.0 | 0.0 | 2 |
| Imexon | Acyclovir | RRM2;GUK1, RRM1;GUK1 | 10.0 | 0.0 | 2 |
| Tioguanine | Pemetrexed | HPRT1;GART, ATIC;HPRT1 | 5.0 | 0.0 | 2 |
| DB01632 | Pemetrexed | HPRT1;GART, ATIC;HPRT1 | 5.0 | 0.0 | 2 |
| Azathioprine | Mercaptopurine | PPAT;HPRT1, SLC29A2;SLC29A1 | 3.5 | 1.0 | 2 |
| Mercaptopurine | Mercaptopurine | PPAT;HPRT1, SLC29A2;SLC29A1 | 3.5 | 1.0 | 2 |
| Human calcitonin | Cystine | ANPEP;SLC3A1, ANPEP;SLC7A9 | 3.0 | 1.0 | 2 |
| Lamivudine | Choline salicylate | CMPK1;SLC22A5, CMPK1;PLD2 | 3.0 | 1.0 | 2 |
| Lamivudine | Choline | CMPK1;SLC22A5, CMPK1;PLD2 | 3.0 | 1.0 | 2 |
| Icatibant | Cystine | ANPEP;SLC3A1, ANPEP;SLC7A9 | 3.0 | 1.0 | 2 |
| Gemcitabine | Choline salicylate | CMPK1;SLC22A5, CMPK1;PLD2 | 3.0 | 1.0 | 2 |
| Gemcitabine | Choline | CMPK1;SLC22A5, CMPK1;PLD2 | 3.0 | 1.0 | 2 |
| Ezetimibe | Cystine | ANPEP;SLC3A1, ANPEP;SLC7A9 | 3.0 | 1.0 | 2 |
| Sofosbuvir | Choline | CMPK1;SLC22A5, CMPK1;PLD2 | 3.0 | 1.0 | 2 |
| Sofosbuvir | Choline salicylate | CMPK1;SLC22A5, CMPK1;PLD2 | 3.0 | 1.0 | 2 |
| Gemcitabine | Phosphatidyl serine | SLC29A2;PTDSS1, CMPK1;PTDSS1 | 2.5 | 2.0 | 2 |
| Taurocholic acid | Phosphatidyl serine | SLC16A1;PTDSS1 | 3.0 | 3.0 | 1 |
| Nateglinide | Phosphatidyl serine | SLC16A1;PTDSS1 | 3.0 | 3.0 | 1 |
| Lamivudine | Phosphatidyl serine | CMPK1;PTDSS1 | 3.0 | 2.0 | 1 |
| Quercetin | Phosphatidyl serine | SLC16A1;PTDSS1 | 3.0 | 3.0 | 1 |
| Probenecid | Phosphatidyl serine | SLC16A1;PTDSS1 | 3.0 | 3.0 | 1 |
| Aminohippuric acid | Phosphatidyl serine | SLC16A1;PTDSS1 | 3.0 | 3.0 | 1 |
| Acetic acid | Phosphatidyl serine | SLC16A1;PTDSS1 | 3.0 | 3.0 | 1 |
| Lactic acid | Phosphatidyl serine | SLC16A1;PTDSS1 | 3.0 | 3.0 | 1 |
| Benzoic acid | Phosphatidyl serine | SLC16A1;PTDSS1 | 3.0 | 3.0 | 1 |
| Ampicillin | Phosphatidyl serine | SLC16A1;PTDSS1 | 3.0 | 3.0 | 1 |
| Salicylic acid | Phosphatidyl serine | SLC16A1;PTDSS1 | 3.0 | 3.0 | 1 |
| Arbaclofen Placarbil | Phosphatidyl serine | SLC16A1;PTDSS1 | 3.0 | 3.0 | 1 |
| Methotrexate | Phosphatidyl serine | SLC16A1;PTDSS1 | 3.0 | 3.0 | 1 |
| Valproic acid | Phosphatidyl serine | SLC16A1;PTDSS1 | 3.0 | 3.0 | 1 |
| Foscarnet | Phosphatidyl serine | SLC16A1;PTDSS1 | 3.0 | 3.0 | 1 |
| Sofosbuvir | Phosphatidyl serine | CMPK1;PTDSS1 | 3.0 | 2.0 | 1 |
| Pravastatin | Phosphatidyl serine | SLC16A1;PTDSS1 | 3.0 | 3.0 | 1 |
| gamma-Hydroxybutyric acid | Phosphatidyl serine | SLC16A1;PTDSS1 | 3.0 | 3.0 | 1 |